\documentclass[%
aps,
prx,
reprint,
superscriptaddress,
amsmath,
amssymb,
longbibliography
]{revtex4-1}
\usepackage{
    physics,
    graphicx,
    booktabs,
    mathtools, %for \begin{rcases} \end{rcases}
    placeins, %To include \FloatBarrier
    hyperref %To use put in url, enable links in bibiliography
}
\usepackage[dvipsnames]{xcolor}
\hypersetup{
    colorlinks,
    linkcolor={blue!50!black},
    citecolor={blue!50!black},
    urlcolor={blue!80!black}
}% remove the box around citation link and bibilio link

\usepackage[caption=false]{subfig} %The caption part is not compatible with revtex
\usepackage{tikz}
\usetikzlibrary{quantikz}

\newcommand{\supop}[1]{\ensuremath{\overbracket[0.1ex][0.3ex]{#1}}}
\newcommand{\supsupop}[1]{\ensuremath{\overbracket[0.1ex][0.2ex]{\overbracket[0.1ex][0.2ex]{#1}}}}
\DeclarePairedDelimiter\pbra{\langle\!\langle}{\rvert}
\DeclarePairedDelimiter\pket{\lvert}{\rangle\!\rangle}
\DeclarePairedDelimiterX\pbraket[2]{\langle\!\langle}{\rangle\!\rangle}{#1 \delimsize\vert #2}
\newcommand{\addressOxMat}{\affiliation{Department of Materials, University of Oxford, Oxford, OX1 3PH, United Kingdom}}
\newcommand{\addressQMT}{\affiliation{Quantum Motion Technologies Ltd, Nexus, Discovery Way, Leeds, West Yorkshire, LS2 3AA, United Kingdom}}

\begin{document}
    
    %\preprint{APS/123-QED}
    
    \title{Mitigating Coherent Noise Using Pauli Conjugation}% Force line breaks with \\
    
    \author{Zhenyu Cai} \email{zhenyu.cai@materials.ox.ac.uk} \addressOxMat\addressQMT
    \author{Xiaosi Xu} \addressOxMat
    \author{Simon C. Benjamin} 
    \email{simon.benjamin@materials.ox.ac.uk}
    \addressOxMat\addressQMT
    
    \date{\today}% It is always \today, today,
    %  but any date may be explicitly specified
    
    \begin{abstract}
        Coherent noise can be much more damaging than incoherent (probabilistic) noise in the context of quantum error correction. One solution is to use twirling to turn coherent noise into incoherent Pauli channels. In this Article, we show that some of the coherence of the noise channel can actually be used to improve its logical fidelity by simply sandwiching the noise with a chosen pair of Pauli gates, which we call Pauli conjugation. Using the optimal Pauli conjugation, we can achieve a higher logical fidelity than using twirling and doing nothing. We devise a way to search for the optimal Pauli conjugation scheme and apply it to Steane code, 9-qubit Shor code and distance-3 surface code under global coherent $Z$ noise. The optimal conjugation schemes show improvement in logical fidelity over twirling while the weights of the conjugation gates we need to apply are lower than the average weight of the twirling gates. In our example noise and codes, the concatenated threshold obtained using conjugation is consistently higher than the twirling threshold and can be up to 1.5 times higher than the original threshold where no mitigation is applied. Our simulations show that Pauli conjugation can be robust against gate errors. With the help of logical twirling, the undesirable coherence in the noise channel can be removed and the advantages of conjugation over twirling can persist as we go to multiple rounds of quantum error correction. 
        \\
        \\
        \\
        \\
    \end{abstract}
    
    \maketitle
    \section{Introduction}
    The quantum fault-tolerant threshold theorem states that when the error rate of the physical components is below a certain threshold value for a given quantum error correction code, we can reduce the error rate of the logical qubits indefinitely by scaling up our code~\cite{aharonovFaulttolerantQuantumComputation1997,knillResilientQuantumComputation1998,aliferisQuantumAccuracyThreshold2006}. Thus for a given code, its threshold value is the target hardware error rate the experimentalists will aim for. The threshold error rate is defined using the worst case error rate like \emph{the diamond distance} since it is related to the rate of error accumulation. However, experimentally we can only measure the average error rate like \emph{the fidelity} efficiently. For Pauli channels, the worst case error rate is similar to the average case error rate. However, for coherent (unitary) errors, their worst case error rate can scale as the square root of the average error rate, making them potentially more damaging to quantum error correction codes due to a faster rate of error accumulation~\cite{sandersBoundingQuantumGate2015, gutierrezComparisonQuantumErrorcorrection2015, kuengComparingExperimentsFaultTolerance2016a, bravyiCorrectingCoherentErrors2018, greenbaumModelingCoherentErrors2018,iyerSmallQuantumComputer2018, huangPerformanceQuantumError2019}.
    
    At the physical qubit level, coherent noise can be mitigated using dynamical decoupling~\cite{lidarReviewDecoherenceFree2014, suterColloquiumProtectingQuantum2016}, however there are limitations due to imperfect control pulses and finite pulse durations and intervals. In the context of quantum error correction, local physical coherent noise will be decohered at the logical level as the code scale up~\cite{bealeQuantumErrorCorrection2018}. Their damage to the encoded state can be mitigated by using better decoders~\cite{chamberlandHardDecodingAlgorithm2017}. Gate-level coherent errors in quantum error correction circuit can be mitigated by splitting the stabiliser check into two oppositely rotating halves~\cite{debroyStabilizerSlicingCoherent2018} with some requirements on the gates available to the given architecture. A more general solution would involve using Pauli twirling to turn the coherent noise into a Pauli channel~\cite{bennettPurificationNoisyEntanglement1996, bennettMixedstateEntanglementQuantum1996, knillRandomizedBenchmarkingQuantum2008, emersonSymmetrizedCharacterizationNoisy2007}, which as mentioned before can be much less damaging to the fault-tolerant threshold. Twirling generally involves using all possible Pauli gates to sandwich the noise channel and averaging over the results. The average weight of the extra twirling gates we need to apply scales with the total number of qubits, thus the gate errors introduced by the twirling gates are not negligible. 
    
    In this Article, instead of using twirling to combat coherent errors, we propose to deterministically sandwich the noise channel using a chosen pair of Pauli gates, which we call Pauli conjugation. We start by introducing some background concepts in Section~\ref{sect:background}. In Section~\ref{sect:find_opt_conj}, we find ways to reduce the search space for the optimal Pauli conjugation scheme. This is then used in Section~\ref{sect:Z_noise_sim} to compare the logical fidelity and concatenated threshold of Pauli conjugation to those of twirling for several quantum error correction codes under global $Z$ rotation noise. In Section~\ref{sect:multi}, we discuss the extension of our technique to multiple rounds of error corrections and conjugations. This is followed by conclusion and discussion of possible future directions in Section~\ref{sect:conclusion}.
    \quad\\
    \quad\\
    \quad\\
    \quad\\
    \quad\\
    \section{Logical Fidelity in Quantum Error Correction}\label{sect:background}
    \subsection{Pauli Transfer Matrix Formalism}
    In the Pauli transfer matrix formalism~\cite{greenbaumIntroductionQuantumGate2015}, the density operators are written in vector form by decomposing into Pauli basis $G \in \mathbb{G}$:
    \begin{align*}
    \rho &= \frac{1}{2^n}\sum_{G \in \mathbb{G}} \Tr(G\rho) G \\
    \Rightarrow \pket{\rho} &= \sum_{G \in \mathbb{G}} \pket{G}\pbraket{G}{\rho}
    \end{align*}
    where we have defined the inner product as:
    \begin{align*}
    \pbraket{G}{\rho}  = \frac{1}{\sqrt{2^n}} \Tr(G\rho)
    \end{align*}
    We have added a scaling factor $\frac{1}{\sqrt{2^n}}$ when we use the Pauli operators as basis, where $n$ is the number of qubits. This is to ensure the normalisation of the basis set $\{\pket{G}\}$.
    
    In such a way, a general quantum channel $\mathcal{E}$ can be written in matrix form:
    \begin{align*}
    \mathcal{E} = \sum_{G, G' \in \mathbb{G}} \pket{G'}\pbra{G'}\mathcal{E}\pket{G}\pbra{G}
    \end{align*}
    with the matrix elements given by
    \begin{align*}
    \mathcal{E}_{G'G} = \pbra{G'}\mathcal{E}\pket{G} = \pbraket{G'}{\mathcal{E}(G)} = \frac{1}{2^n}\Tr(G' \mathcal{E}(G)).
    \end{align*}
    
    \subsection{Quantum Error Correction}
    For a code defined by the set of stabilisers $\mathbb{S}$, we will denote the stabiliser generators as $\widetilde{\mathbb{S}}$.
    In this Article, the generator of a set is denoted using $\widetilde{\quad}$. When we talk about the generators for a Pauli set, the composition operation we used in the generation will ignore all the phase factors in front. 
    
    We will do stabiliser measurements for all $\widetilde{S}_i \in \widetilde{\mathbb{S}}$ to extract the error syndrome $\vec{m}$ whose element $m_i \in \{0, 1\}$ is the measurement outcome of the stabiliser generator $\widetilde{S}_i$. This will project the noisy state into the corresponding $\vec{m}$-syndrome subspace using the syndrome projection operators
    \begin{align*}
    \Pi_{\vec{m}} =  \prod_{i = 1}^{\abs{\widetilde{\mathbb{S}}}} \frac{\left(1 + (-1)^{m_i}\widetilde{S}_i\right)}{2}.
    \end{align*}
    For each measured syndrome $\vec{m}$, we will apply the corresponding recovery operator $R_{\vec{m}}$, which is usually chosen to be the most likely Pauli error that leads to the given syndrome. Using $\supop{\quad}$ to denote a super-operator
    \begin{align*}
    \left(\supop{A} + \supop{B}\right) \rho = A \rho A^\dagger + B \rho B^\dagger,
    \end{align*}
    the overall quantum error correction process can be written as:
    \begin{align*}
    \mathcal{C} = \sum_{\vec{m}} \supop{R_{\vec{m}}\Pi_{\vec{m}}} = \sum_{\vec{m}} \supop{\Pi_{0} R_{\vec{m}}}
    \end{align*}
    where we have used $\Pi_{\vec{m}} = R_{\vec{m}} \Pi_{0} R_{\vec{m}}$.
    
    If we start within the logical subspace, the error correction process $\mathcal{C}$ will always project the state back to the logical subspace even after going through a noisy channel $\mathcal{N}$. Thus, the effective channel $\mathcal{N}_{0} = \mathcal{C}\mathcal{N}$ will be a error channel that takes one logical state to another, i.e. it is a logical noise channel. The effective logical noise channel $\overline{\mathcal{N}}_{0}$ is defined to be the average over all logically equivalent starting and final states:
    \begin{align}
    \overline{\mathcal{N}}_{0} & = \sum_{G, G' \in \mathbb{G}} \pket{\overline{G}'\Pi_{0}}\pbra{\overline{G}'\Pi_{0}} \mathcal{C} \mathcal{N} \pket{\overline{G}\Pi_{0}}\pbra{\overline{G}\Pi_{0}}\nonumber\\
    & = \sum_{\vec{m}} \sum_{G, G' \in \mathbb{G}} \pket{\overline{G}'\Pi_{0}}\pbra{\overline{G}'\Pi_{0}} \supop{\Pi_{0} R_{\vec{m}}} \mathcal{N} \pket{\overline{G}\Pi_{0}}\pbra{\overline{G}\Pi_{0}}\nonumber\\
    & = \sum_{\vec{m}} \sum_{G, G' \in \mathbb{G}} \pket{\overline{G}'\Pi_{0}}\pbra{\overline{G}'\Pi_{0}} {\supop{R}}_{\vec{m}} \mathcal{N} \pket{\overline{G}\Pi_{0}}\pbra{\overline{G}\Pi_{0}}\nonumber\\
    & = \mathcal{R} \mathcal{N} \label{eqn:eff_log_channel}
    \end{align}
    where $\mathcal{R} = \sum_{\vec{m}} {\supop{R}}_{\vec{m}}$~\footnote{Note that here we have abused the notation of $\mathcal{R} \mathcal{N}$ assuming it will only act on the logical Pauli basis $\{\pket{\overline{G}\Pi_{0}}\}$ instead on all of the physical Pauli basis.}. 
    
    \subsection{Twirling and Conjugation}\label{sect:twirling}
    Twirling is a technique for converting an arbitrary error channel into a Pauli channel~\cite{bennettMixedstateEntanglementQuantum1996, bennettPurificationNoisyEntanglement1996, gellerEfficientErrorModels2013}, which is carried out by taking the average of the error channel conjugated with different gates chosen from a set of Pauli gates $\mathbb{W} \subseteq \mathbb{G}$ that we call the twirling set. Conventionally, twirling is carried out using the full set of Pauli gates as the twirling set: $\mathbb{W} = \mathbb{G}$. However, it is possible to find a smaller $\mathbb{W}$ that is equivalent to the full Pauli set as we will see later (also shown in~\cite{caiConstructingSmallerPauli2019}).
    
    Twirling a noise channel $\mathcal{N}$ is just
    \begin{align}\label{eqn:twirling}
    \mathcal{T}(\mathcal{N}) &= \frac{1}{\abs{\mathbb{W}}}\sum_{W \in \mathbb{W}} \supop{W} \mathcal{N} \supop{W}
    \end{align}
    Twirling can decohere the Pauli components in the noise channel and turn it into a Pauli channel. This will correspond to removing the off-diagonal elements of the Pauli transfer matrix of the channel.
    
    Using (\ref{eqn:eff_log_channel}) and (\ref{eqn:twirling}), the effective logical channel after twirling is:
    \begin{align*}
    \overline{\mathcal{N}}_{T} & = \mathcal{R} \mathcal{T}(\mathcal{N}) \\
    & = \frac{1}{\abs{\mathbb{W}}} \sum_{W \in \mathbb{W}}  \mathcal{R} \supop{W}\mathcal{N}\supop{W}.
    \end{align*}
    
    Instead of averaging over all twirling gates, if we deterministically conjugate the noise process with a given twirling gate $W$, the effective logical channel can be written as
    \begin{align*}
    \overline{\mathcal{N}}(W) = \mathcal{R} \supop{W}\mathcal{N}\supop{W}
    \end{align*}
    which we will call Pauli conjugation.
    
    Then we have:
    \begin{align*}
    \overline{\mathcal{N}}_0 & = \overline{\mathcal{N}}(I) \\
    \overline{\mathcal{N}}_{T} & = \frac{1}{\abs{\mathbb{W}}} \sum_{W \in \mathbb{W}}  \overline{\mathcal{N}}(W).
    \end{align*}
    
    The logical fidelity of $\overline{\mathcal{N}}(W)$ is%~\cite{magesanGainingInformationQuantum2008}
    \begin{align*}
    F(W) &= \int \pbra{\overline{\rho}}  \overline{\mathcal{N}}(W)\pket{\overline{\rho}} \dd{\overline{\rho}}  
    \end{align*}
    where $\overline{\rho}$ is a logical state and the integral is over the \emph{pure state} surface using the Haar measure.
    
    Since the fidelity $F$ is a linear function of the noise process $\overline{\mathcal{N}}$, we can similarly obtain the original logical fidelity $F_0$ and the twirled logical fidelity $F_T$:
    \begin{align*}
    F_0 &= F(I)\\
    F_T &= \frac{1}{\abs{\mathbb{W}}} \sum_{W \in \mathbb{W}} F(W) .
    \end{align*}
    
    There exists a $W_{max} \in \mathbb{W}$ such that $F(W_{max})$ is the maximum $F(W)$ that we can achieve. By definition we have
    \begin{align*}
    F(W_{max}) &\geq F(I)\\
    F(W_{max}) &\geq F_{T}.
    \end{align*}
    Thus if we can find such $W_{max}$ and deterministically apply it to the noise instead of doing nothing or randomly applying all $W \in \mathbb{W}$, we can obtain a higher fidelity $F(W_{max})$ than the original fidelity $F(I)$ and the twirled fidelity $F_{T}$. 
   
    \subsection{Mechanism of Conjugation}\label{sect:mech_conj}
    Let us first consider the case when we perform quantum error correction on a unitary (completely coherent) noise channel and obtain the $0$-syndrome ($m_i = 0 \quad \forall i$). The resultant effective noise channel will contain an error-free components representing by the coherent superposition of the stabiliser operators $\sum_i \alpha_i S_i$. When acting on a logical state, the effective amplitude corresponding to the logical identity will then be $\sum_i \alpha_i$. 
    
    Now if we apply Pauli conjugation using the operator $W$ to the error channel, the error-free components will become $\sum_i \alpha_i WS_iW$, which corresponds to an amplitude of $\sum_i \eta(W, S_i)\alpha_i$ for the logical identity. Here $\eta(A, B)$ is the commutator between operators $A$ and $B$:
    \begin{align*}
    AB = \eta(A,B) BA.
    \end{align*}
    
    Thus Pauli conjugation will change the sign of the Pauli components of the error channel, changing the way the Pauli components interfere. For the $0$-syndrome case, if we can choose a conjugation operator $W$ such that $\sum_i \eta(W, S_i)\alpha_i \geq \sum_i \alpha_i$, i.e. the error-free components (the stabilisers) interfere more constructively with conjugation than without, it will lead to an increase in the logical fidelity of the channel using conjugation. The normalisation of the channel also means that the logical error components of the channel will interfere more destructively when using conjugation. Similar arguments can be made for the non-zero-syndrome cases.
    
    Hence for a given noise channel, as long as there is some coherent superposition of its Pauli components corresponding to the same logical operators for a given syndrome, Pauli conjugation should be able to improve its logical fidelity by changing the relative signs between the components and alter the way they interfere. One case for which Pauli conjugation will not be able to help is when the identity is the optimal conjugation gate $W_{max} = I$, i.e. the noise Pauli components are interfering in the optimal ways for the given code, which should be unlikely unless we have hand-picked our code to exactly fit the noise process.

    \section{Finding the optimal conjugation gate}\label{sect:find_opt_conj}
    The usual Pauli twirling will have $\mathbb{W} = \mathbb{G}$. For $n$ qubits, this means that there are $4^n$ elements in $\mathbb{W}$ that we need to search over to find $W_{max}$, which is exponentially difficult for large $n$. Hence, we first need to reduce the size of $\mathbb{W}$ in order to find $W_{max}$ effectively. 
    
    Rather than dealing with the twirling set $\mathbb{W}$, we will first be working with its generator $\widetilde{\mathbb{W}}$. The reason we can work with the generators for our later purposes is outlined in Appendix~\ref{sect:generator_reduce}. 
    
    The generators of the conventional twirling set is just $\widetilde{\mathbb{W}} = \widetilde{\mathbb{G}}$. For a given quantum error correction code, the generators of the Pauli basis $\widetilde{\mathbb{G}}$ can be divided into the following partitions:
    \begin{itemize}
        \item Stabiliser generators $\widetilde{\mathbb{S}}$: the set of Pauli operators that define the stabiliser checks of the code.
        \item Logical generators $\widetilde{\overline{\mathbb{G}}}$: together with the stabiliser generators, they generate the set of logical operators $\overline{\mathbb{G}}$, which is just the normaliser of the set of stabilisers $\mathbb{S}$.
        \item Error generators $\widetilde{\mathbb{E}}$: All the remaining generators needed to generate the whole Pauli set. Each error generator $\widetilde{E}$ anti-commutes with a different subset of stabiliser generators and thus will produce a different syndrome. 
    \end{itemize}
    Hence, we have
    \begin{align*}
    \widetilde{\mathbb{W}} = \widetilde{\mathbb{G}} = \widetilde{\mathbb{S}} + \widetilde{\mathbb{E}} + \widetilde{\overline{\mathbb{G}}}
    \end{align*}
    Note that we have used the label `error generators' since each such element creates a code violation, but physical error process can give rise to elements of any of these sets, and in particular those in $\widetilde{\overline{\mathbb{G}}}$ which create undetectable logical errors.
    
    \subsection{Removing Stabilisers and Logical Operators}\label{sect:remove_stb}
    $\mathcal{R}$ and $\supop{S}$ commute because they are both Pauli channels which are diagonal in the form of Pauli transfer matrix. Hence, for any channel $\mathcal{N}$, and logical states $\pket{\overline{\rho}}$ and $\pket{\overline{\rho}'}$, we have:
    \begin{align*}
    \pbra{\overline{\rho}'} \mathcal{R} \supop{S}\mathcal{N}\supop{S}\pket{\overline{\rho}} & = \pbra{\overline{\rho}'} \supop{S} \mathcal{R} \mathcal{N}\supop{S}\pket{\overline{\rho}} = \pbra{\overline{\rho}'} \mathcal{R} \mathcal{N} \pket{\overline{\rho}}
    \end{align*}
    which means that conjugation using stabilisers on any noise channel has a trivial effect on the effective logical channels. Hence, we can remove all stabilisers from the twirling generator set and reduce it to:
    \begin{align*}
    \widetilde{\mathbb{W}} = \widetilde{\mathbb{E}} + \widetilde{\overline{\mathbb{G}}}
    \end{align*}
    Now if we are calculating the logical fidelity, we are integrating over all the logical pure state using the unitary Haar measure, which is by definition invariant under any unitary transformation. Thus we have:
    \begin{align*}
    F_0 & = \int \pbra{\overline{\rho}}  \mathcal{R} \mathcal{N} \pket{\overline{\rho}} \dd{\overline{\rho}}\\  
    & = \int \pbra{\overline{\rho}} \supop{\overline{G}}  \mathcal{R} \mathcal{N} \supop{\overline{G}}\pket{\overline{\rho}} \dd{\overline{\rho}}
    \end{align*}
    $\mathcal{R}$ and $\supop{\overline{G}}$ again commute since they are both Pauli channel. Hence we have:
    \begin{align*}
    F_0 & =  \int \pbra{\overline{\rho}}   \mathcal{R} \supop{\overline{G}} \mathcal{N} \supop{\overline{G}}\pket{\overline{\rho}} \dd{\overline{\rho}} \\
    & =F(\overline{G}) \quad \forall G \in \mathbb{G}
    \end{align*}
    Hence, when calculating the logical fidelity, conjugation with logical Pauli operators also acts trivially and can be removed from the twirling generating group. The remaining non-trivial twirling generators are:
    \begin{align*}
    \widetilde{\mathbb{W}} = \widetilde{\mathbb{E}} 
    \end{align*}
    The way to construct a $\widetilde{\mathbb{E}}$ consists of only single-qubit $X$ and $Z$ gates is outlined in Appendix~\ref{sect:construct_E}.
    
    \subsection{Twirling Set Reduction Using the Structure of the Noise}\label{sect:noise_sym}
    Two super-operators $\supop{A}$ and $\supop{B}$ will commute if their commutator $\eta(A, B) = e^{i\phi}$, i.e. their commutator is some phase factor. 
    
    We will write our noise channel $\mathcal{N}$ in terms of its noise elements $\supop{N}$:
    \begin{align*}
    \mathcal{N} = \sum_{N} \supop{N}.
    \end{align*}
    Now if a twirling generator $W$ satisfies $\eta(W, N) = e^{i\phi}\ \forall N$, then
    \begin{align*}
    \supop{W}\supop{N}\supop{W} &= \supop{N} \quad \forall N\\
    \supop{W}\mathcal{N}\supop{W} &= \mathcal{N}
    \end{align*}
    i.e. it act trivially on noise $\mathcal{N}$ and hence can be removed.
    
    After such reduction, the twirling generating set now becomes:
    \begin{align*}
    \widetilde{\mathbb{W}} = \{W \in \widetilde{\mathbb{E}}\ |\ \exists N\ \eta(W, N) \neq e^{i\phi} , \phi \in \mathbb{R}\}
    \end{align*}
    
    \subsection{Symmetry in Code and Noise}\label{sect:perm_sym}
    The twirling set $\mathbb{W}$ can be generated from $\widetilde{\mathbb{W}}$ following Appendix~\ref{sect:W_construction}. Based on the symmetry existing in both the code and the noise, we can prove the equivalence between different elements in $\mathbb{W}$.
    
    Suppose we manage to find a Clifford operation $U$ such that the code state basis $\Pi_{\vec{0}}\overline{G}$ and the physical noise channel $\mathcal{N}$ are invariant under its transformation:
    \begin{equation}\label{eqn:symmetry_condition}
    \begin{split}
    \left[U, \Pi_{\vec{0}}\overline{G}\right] &= 0 \quad \forall G \in \mathbb{G}\\
    \left[\supop{U}, \mathcal{N}\right] &= 0
    \end{split}
    \end{equation}
    we can prove that (see Appendix~\ref{sect:perm_sym_deriv})
    \begin{align}\label{eqn:equiv_symmetry} 
    \overline{\mathcal{N}}(W) & = \overline{\mathcal{N}}(U^\dagger W U)
    \end{align}
    i.e. the effective logical channel conjugated with $W$ is the same as that with $U^\dagger W U$. All of such $U$ will form a group $\mathbb{U}$.
    
    Hence, we can define an equivalence relation:
    \begin{align*}
    W' \sim W  \iff \exists U \in \mathbb{U} \quad W' = U^\dagger W U
    \end{align*}
    In such a way, conjugacy with elements in $\mathbb{U}$ will split $\mathbb{W}$ into several equivalence classes. The elements in the same equivalence class will produce the same logical fidelity when used to conjugate the noise.
    
    The simplest type of Clifford transformation to consider is qubit permutation, for which $U$ consists of swap gates. Permutation symmetry of quantum error correction codes has been studied in \cite{huangPerformanceQuantumError2019} and \cite{chamberlandHardDecodingAlgorithm2017}. Note that qubit permutation will preserve the weights of the operators, thus it is crucial to construct $\mathbb{W}$ to have the elements with the lowest weight possible (see Appendix~\ref{sect:W_construction}), so that more of them can be proven to be in the same equivalence class.

    If a code has one logical qubit and its logical Pauli gates consist of applying physical Pauli gates to all the qubits, then such transversal logical Pauli gates are invariant~\footnote{More precisely, we can find at least one physical representation of logical $\overline{G}$ out of all of its logically equivalent counter-parts that satisfy this symmetry condition.} under any qubit permutation $U$, i.e. $\left[U, \overline{G}\right] = 0 \quad \forall G \in \mathbb{G}$. For such codes, we only need to further make sure that the set of stabilisers are invariant under the given qubit permutation $U$:
    \begin{align}\label{eqn:stb_sym}
    \left[U, \Pi_{0}\right] = 0
    \end{align}
    to ensure the code symmetry requirement in (\ref{eqn:symmetry_condition}) is satisfied. Furthermore, if some of the stabilisers commute with the noise, then these stabilisers will have trivial effect in the error correction process and thus can be safely ignored. In such a case, we will only need to consider the symmetry of the stabilisers that do not commute with the noise. For example, for a pure $Z$ noise, we can safely ignore the $Z$ stabilisers when we are considering code symmetry.
    
    \section{Mitigating coherent $Z$ noise using Pauli conjugation}\label{sect:Z_noise_sim}
    In this Section we will try to find the optimal Pauli conjugation gate for different quantum error correction codes under the global $Z$ rotation noise:
    \begin{align}
    N(\theta) &= \prod_{j = 1}^{J} e^{-i\theta Z_j} \label{eqn:noise_model}
    \end{align}
    where $J$ is the number of qubit. This noise is a coherent superposition of all possible $Z$ operators (tensor products of $I$ and $Z$). The weight-$n$ $Z$ operators in the superposition will have the amplitude $\left(-i\sin \theta\right)^n \left(\cos \theta\right)^{J-n}$.
    
    Of course if we are allowed to flip all the qubits right in the middle of the channel, we can flip the direction of the rotation and cancel the coherent error, which is just a simple example of dynamical decoupling. However, if we look at for example high frequency global $Z$ noise, in which the direction of the global $Z$ rotation may flip after a very short time interval in a random walk fashion, dynamical decoupling cannot be applied. In such a case, we have discussed how Pauli conjugation can be used to mitigate such noise in Appendix~\ref{sect:high_freq}. It builds from our discussion in this section, in which we will be looking at the coherent global $Z$ rotation noise described in (\ref{eqn:noise_model}) without allowing gates to be performed in the middle of the channel.
    
    Since the noise only consists of $Z$ components, all pure $Z$ twirling generators will act trivially on the noise, thus can be removed. This noise is symmetric under any qubit permutation. Hence, any permutation symmetry of the quantum error correction code will also exist for the noise.
    
    For all the codes that we will discuss in this section, their logical Pauli gates consist of applying physical Pauli gates to all the qubits. Thus the code symmetry condition in Section~\ref{sect:perm_sym} can be reduced into (\ref{eqn:stb_sym}). Along with the fact that we have pure Z noise, we only need to focus on the symmetry of the $X$ stabilisers in this section when we talk about the symmetry of a code, except for the five-qubit code. Global logical Pauli gates also mean that $N(\frac{\pi}{2})$ will be the $Z$ logical operator. Thus the logical fidelity curve against different $\theta$ will have a rotational symmetry about $\theta = \frac{\pi}{4}$ (see Appendix~\ref{sect:shape_fidelity}), which means that we only need to look at $0\leq \theta \leq \frac{\pi}{4}$ to see the effect of the noise on logical fidelity. 
    
    \subsection{Steane Code}
    \begin{figure}[ht!]
        \centering
        \includegraphics[width=0.25\textwidth]{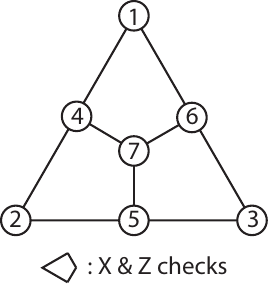}
        \caption{The Steane code.}
        \label{fig:steane_code}
    \end{figure}
    In the Steane Code, we have
    \begin{itemize}
        \item Stabiliser generators $\widetilde{\mathbb{S}}$: $X$ or $Z$ checks on plaquettes $(1,4,6,7)$, $(2,4,5,7)$ and $(3,5,6,7)$.
        \item Logical generators $\widetilde{\overline{\mathbb{G}}}$: $X$ or $Z$ on all qubits.
    \end{itemize}
    Following Section~\ref{sect:remove_stb}, we can construct our twirling generators to be
    \begin{align*}
    \widetilde{\mathbb{W}} = \widetilde{\mathbb{E}} = \{X_1, X_2, X_3, Z_1, Z_2, Z_3\}.
    \end{align*}
    Since the noise only consists of $Z$ components, all pure $Z$ twirling generators will act trivially on the noise, thus can be removed, we then have:
    \begin{align*}
    \widetilde{\mathbb{W}} = \{X_1, X_2, X_3\}
    \end{align*}
    which generates the twirling set:
    \begin{align*}
    \mathbb{W}= \{I, X_1, X_2, X_3, X_4, X_5, X_6, X_7\}.
    \end{align*}
    Note that here we have transformed the error operators to their lowest weight equivalence that produce the same error syndromes.
    
    The Steane code has the same symmetry as the Fano plane~\cite{huangPerformanceQuantumError2019}, whose permutation symmetry group will be denoted as $\mathbb{U}$. Since our noise model is symmetric under any qubit permutation, all $U \in \mathbb{U}$ satisfied (\ref{eqn:symmetry_condition}). 
    
    Now for every pair of single-qubit $X$ operators $X_i, X_j \in \mathbb{W}$, we can find at least one $U \in \mathbb{U}$ such that 
    \begin{align*}
    U^\dagger X_iU = X_j.
    \end{align*}
    Hence, using (\ref{eqn:equiv_symmetry}) we know that all the remaining single-qubit $X$ twirling operators are equivalent. 
    
    There are two equivalence class of twirling gates here, one is equivalent to $I$, while the other is equivalent to $X_1$ (or any single-qubit $X$ gate). 
    
\begin{figure}[htbp!]
        \centering
        \includegraphics[width=0.5\textwidth]{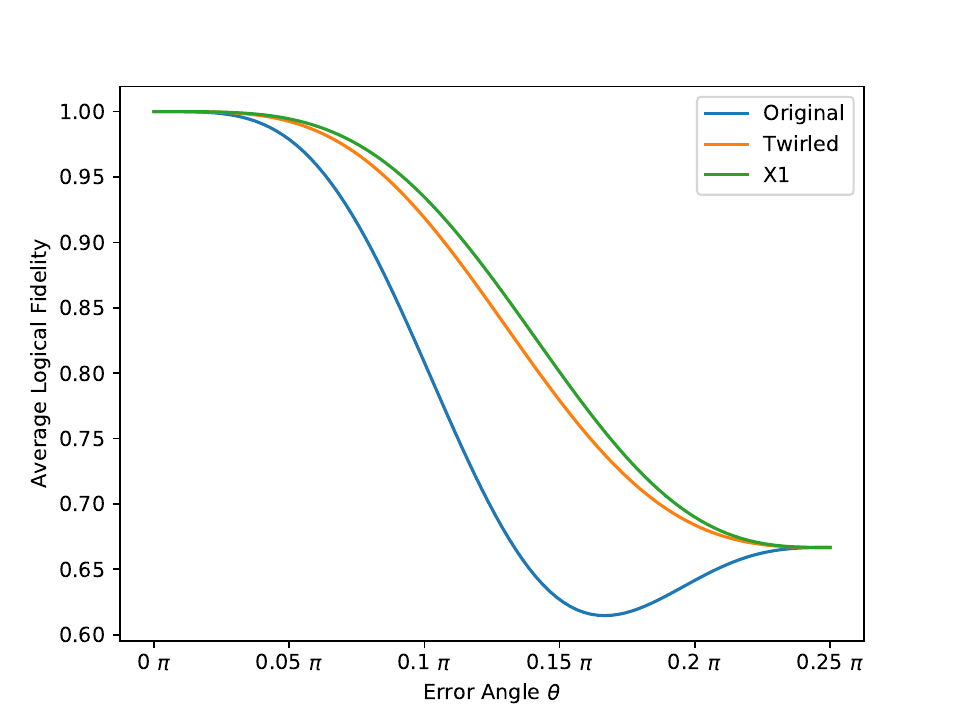}
        \caption{Logical fidelity of the Steane code under different noise strength and noise tailoring schemes.}
        \label{fig:steane_fidelity}
    \end{figure}

    The effect of different strategies on the logical fidelity of Steane code is shown in Figure~\ref{fig:steane_fidelity}. We can see that twirling is consistently better than doing nothing, while $X_1$ conjugation will yield even higher fidelity than twirling.

    \subsection{Other Codes}
    In this section, we will explore the effect of Pauli conjugation using other codes under the same noise model. The details of finding the equivalent class of conjugating gates for different codes are outlined in Appendix~\ref{sect:other_code_details}. Here we will just look at the effect of using conjugating gates in different equivalence classes and compare their effects to doing nothing and twirling.
    
    \subsubsection{Five-qubit code}
\begin{figure}[htbp!]
        \centering
        \includegraphics[width=0.25\textwidth]{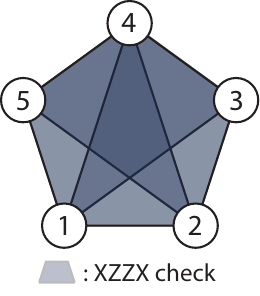}
        \caption{The five-qubit code.}
        \label{fig:five_qubit}
    \end{figure}

    The structure of five-qubit code is shown in Figure~\ref{fig:five_qubit}.
    There is just one non-trivial conjugating strategy in five-qubit code, which is conjugation with any single-qubit X gate, the same we found in the Steane code. However, in our noise model, we found that this strategy makes no difference to the logical fidelity compared to doing nothing. Consequently, the twirled logical fidelity is also the same. Hence, rather interestingly under our noise model, none of the strategies works for the five-qubit code.
    \subsubsection{Nine-qubit Shor code}
\begin{figure}[htbp!]
        \centering
        \includegraphics[width=0.25\textwidth]{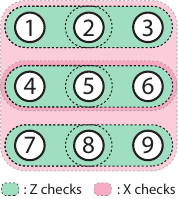}
        \caption{The nine-qubit Shor code.}
        \label{fig:nine_shor}
    \end{figure}

\begin{figure}[htbp!]
        \centering
        \includegraphics[width=0.5\textwidth]{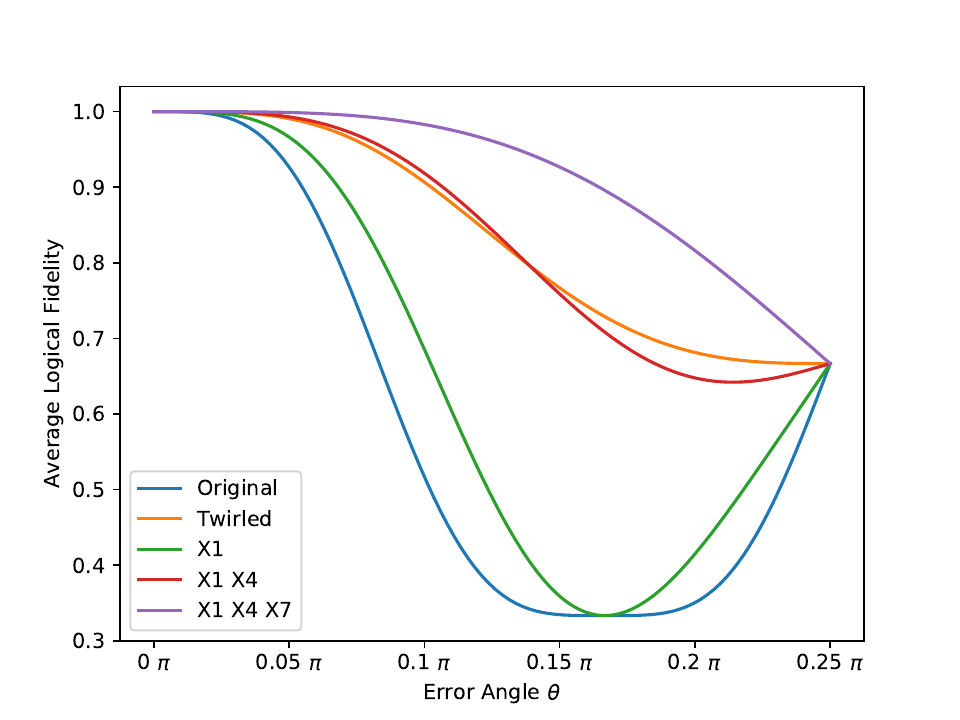}
        \caption{Logical fidelity of the nine-qubit shor code under different noise strength and noise tailoring schemes.}
        \label{fig:shor_code_fidelity}
    \end{figure}

    The structure of the nine-qubit Shor code is shown in Figure~\ref{fig:nine_shor}.
    There are three types of non-trivial Pauli conjugations in the nine-qubit Shor code for our noise model:
    \begin{itemize}
        \item Single qubit flip: $X_1$
        \item Two-qubit flip (in different rows): $X_1X_4$
        \item Three-qubit flip (in different rows): $X_1X_4X_7$
    \end{itemize}
    The effects of these strategies on the logical fidelity are shown in Figure~\ref{fig:shor_code_fidelity}. We see that doing nothing will result in a dip at $\theta = \frac{\pi}{6}$, where our noise turns into a logical operator. Twirling can definitely mitigate such a problem, leading to a great jump in fidelity. Superior improvements can be achieved by conjugating the noise with $X_1X_4X_7$.
    
    The result for the other nine-qubit Shor code with the X and Z checks exchanged is shown in Appendix~\ref{sect:other_shor}.
    \subsubsection{Distance-3 surface code}
\begin{figure}[htbp!]
        \centering
        \includegraphics[width=0.25\textwidth]{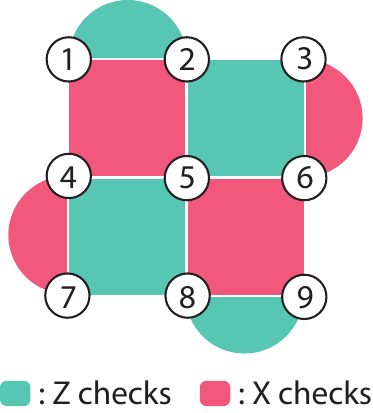}
        \caption{The surface code of distance 3.}
        \label{fig:surface_code}
    \end{figure}

\begin{figure}[htbp!]
        \centering
        \includegraphics[width=0.5\textwidth]{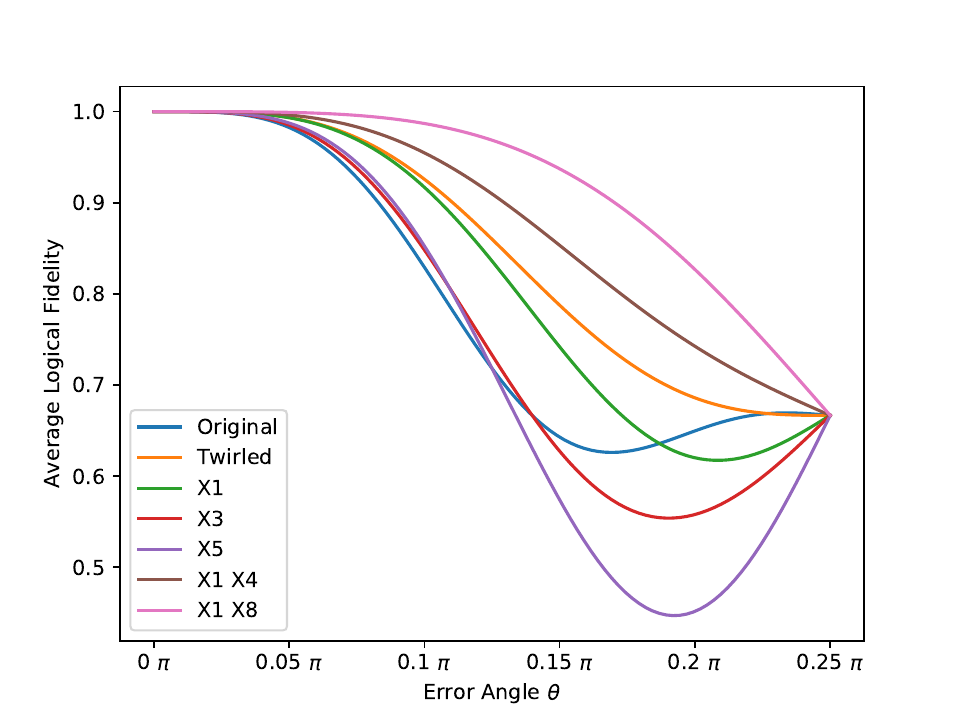}
        \caption{Logical fidelity of the distance-3 surface code under different noise strength and noise tailoring schemes.}
        \label{fig:surface_code_fidelity}
    \end{figure}

    The structure of the distance-3 surface code is shown in Figure~\ref{fig:surface_code}.
    The non-trivial conjugating strategies and their effects on the logical fidelity are shown in Figure~\ref{fig:surface_code_fidelity}. Again we see improvement of the twirled fidelity over doing nothing, and a marked improvement of conjugating the noise with $X_1 X_8$ over twirling.
    
    \subsection{Gate Error}
\begin{figure}[htbp!]
        \centering
        \subfloat[]{\includegraphics[width=0.5\textwidth]{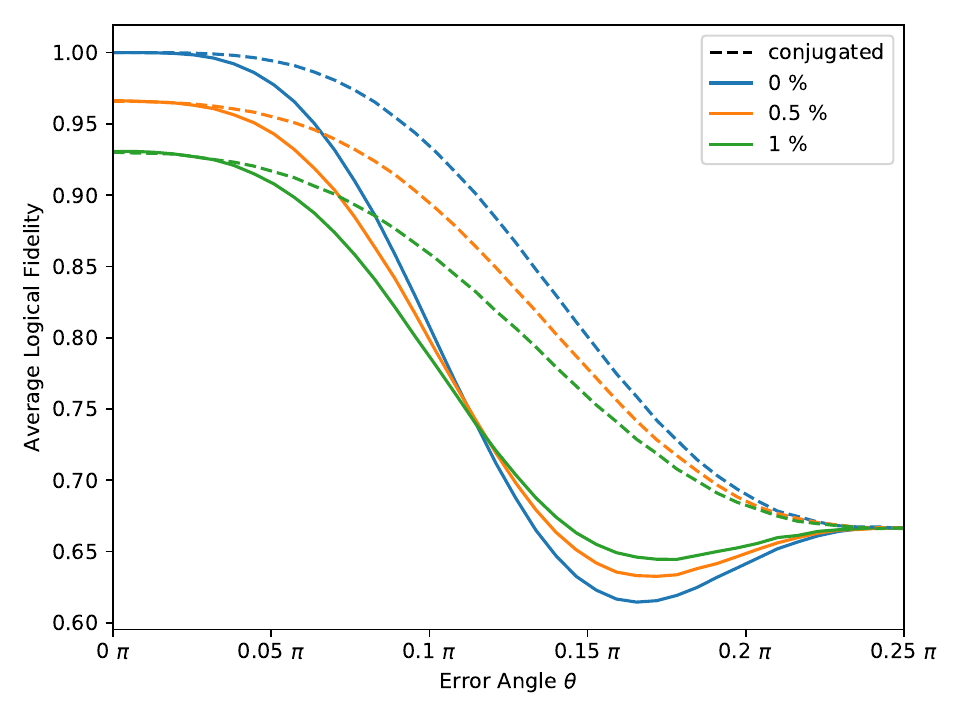}}\\
        \subfloat[]{\includegraphics[width=0.5\textwidth]{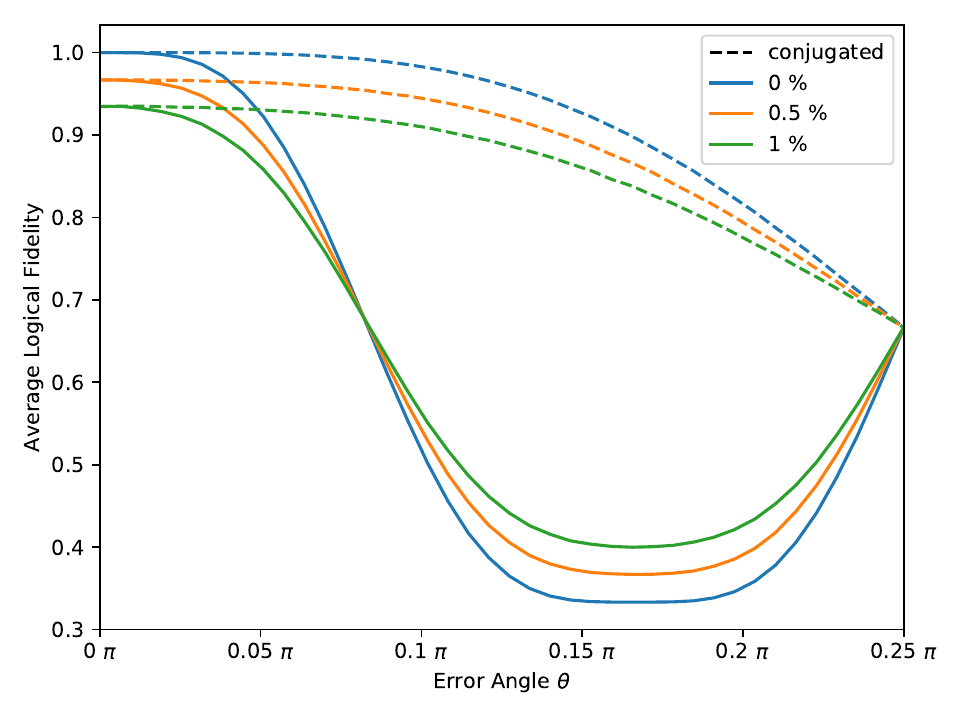}}\\
        \subfloat[]{\includegraphics[width=0.5\textwidth]{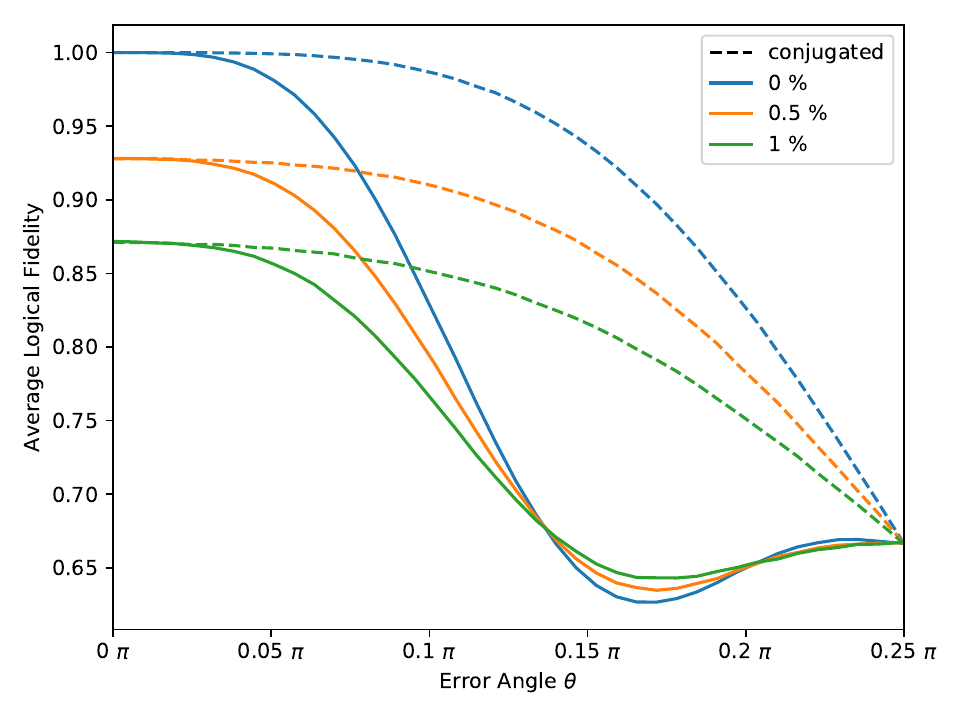}}
        \caption{Logical fidelity with global $Z$ rotation environmental noise of magnitude $\theta$ and depolarising gate noise of probability $0\%$, $0.5\%$ and $1\%$ with or without Pauli conjugation for (a) Steane code, (b) 9-qubit Shor code and (c) distance-3 surface code. The Pauli conjugations we used here are the optimal schemes that we found with zero gate error.}
        \label{fig:FT_fidelity}
    \end{figure}

    In reality, applying extra Pauli gates does not come free due to the errors associated with the gates. We should expect the effect of such errors due to Pauli conjugation to be small since the quantum error correction circuits involve far more gates than Pauli conjugation and also contain two-qubit gates which usually have much lower fidelity than single-qubit Pauli gates. Here we have simulated the performance of different schemes using different codes with depolarising gate error rates of $0.5\%$ and $1\%$ for the encoding circuit, the quantum error correction circuit and the Pauli conjugation gates (with the details of the circuits shown in Appendix~\ref{sect:encoding_circ}). From the result in Figure~\ref{fig:FT_fidelity} we can see that as we increase the gate error rate, the fidelity curves shift downward without much change to their shapes. Hence, the optimal Pauli conjugation schemes maintain their advantages over doing nothing when we take into account gate errors. The fidelity curves using twirling are not shown. However in our examples, we should expect the advantage of Pauli conjugation over twirling increases with increasing gate error rates since the average weights of the twirling gates are higher than that of the conjugation gates. 
    
    When trying to implement Pauli conjugation in practice, such gate errors can be mitigated by absorbing the conjugation gates into the existing gates in the circuit. Such a strategy has been proven to be effective in the case of twirling~\cite{wallmanNoiseTailoringScalable2016}.
    
    \subsection{Concatenated Threshold}
\begin{figure}[htbp!]
        \centering
        \subfloat[]{\includegraphics[width=0.5\textwidth]{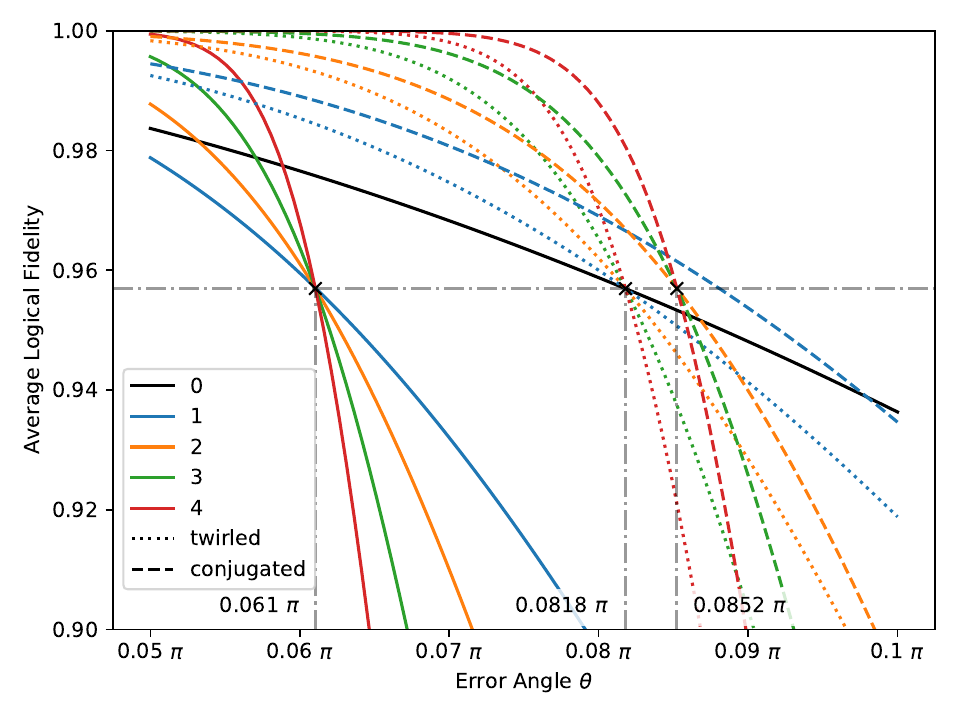}}\\
        \subfloat[]{\includegraphics[width=0.5\textwidth]{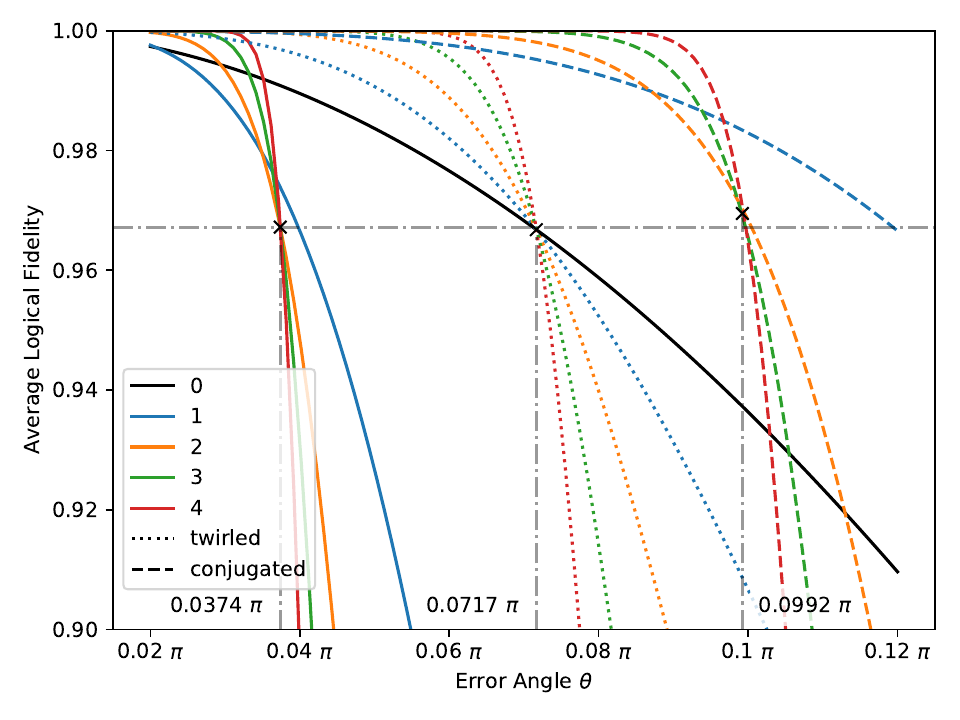}}\\
        \subfloat[]{\includegraphics[width=0.5\textwidth]{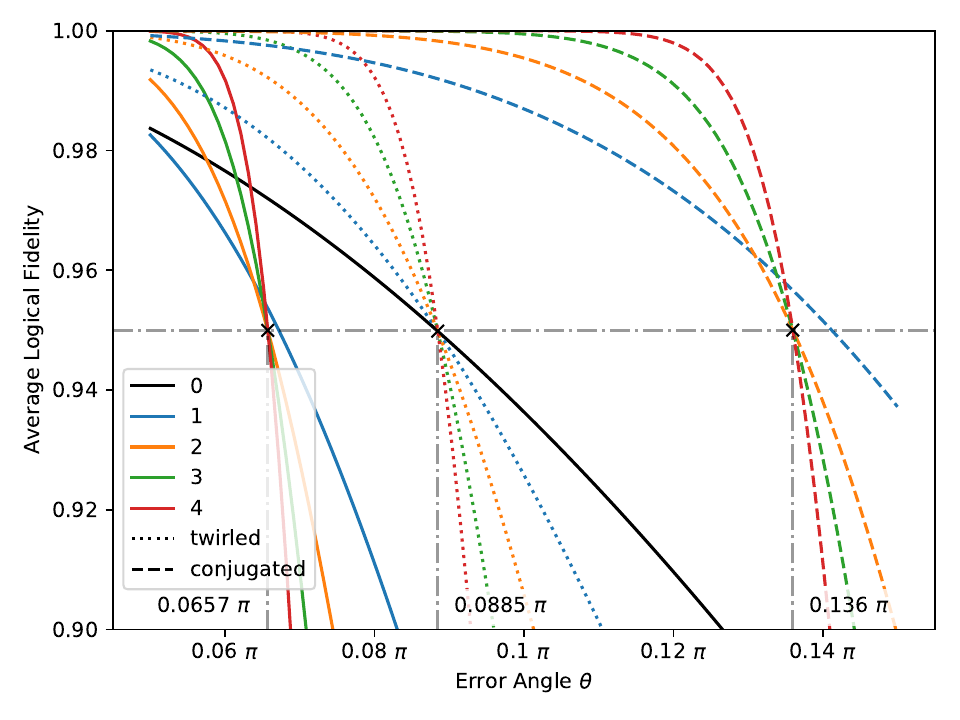}}
        \caption{The concatenated threshold plot under global $Z$ rotation noise for (a) Steane code, (b) 9-qubit Shor code and (c) distance-3 surface code. Different colours shows different levels of concatenation while different line styles shows applying different strategies like twirling or Pauli conjugation to the noise. The Pauli conjugations we used here are the optimal schemes that we found with zero gate error.}
        \label{fig:threshold_plots}
    \end{figure}

    As discussed by Rahn et al. \cite{rahnExactPerformanceConcatenated2002}, after finding the map between the physical noise channel and the logical noise channel with one level of encoding, composing this map will give us the physical-logical noise map for the concatenated code. Here we have assumed that we are using a hard decoder which only takes into account of syndrome information of the current concatenation level. Finding such maps will allow us to compute the performance of a code with different levels of concatenation and hence find its concatenated threshold. Such analysis was carried out in \cite{huangPerformanceQuantumError2019} for a variety of codes. Here we will use the local $Z$ noise map obtained in \cite{huangPerformanceQuantumError2019} to calculate the concatenated threshold for different codes when we apply different kinds of noise tailoring schemes at the physical level (not at any subsequent levels of concatenation). From the results in Figure~\ref{fig:threshold_plots}, we can see the logical fidelity of the threshold crossing points of different noise tailoring schemes are essentially the same. Hence when we try to achieve the threshold logical fidelity with one level of encoding, if one scheme has a higher tolerance of the physical error than another scheme, we should expect a similar improvement in the concatenated threshold. The improvement of the conjugated threshold over the original threshold is $40 \%$, $160\%$ and $110 \%$ for the Steane code, 9-qubit Shor code and distance-3 surface code respectively. All of them also show improvements of the conjugated thresholds over the twirled thresholds.
    
    \section{Composing the Error Channels}\label{sect:multi}
    \subsection{Multiple Rounds of Quantum Error Correction}
    As mentioned in the introduction, coherent errors can be more damaging than incoherent errors because they can accumulate at a faster rate~\cite{sandersBoundingQuantumGate2015, greenbaumModelingCoherentErrors2018}. Hence, we use Pauli twirling to remove the coherent nature of the error channel for each round of error correction so that errors in multiple rounds of error correction will accumulate at a more favourable scaling. As demonstrated in the previous sections, Pauli conjugation can improve the logical fidelity for coherent errors in one round of error correction. However, the error remains coherent after conjugation, which means that the advantages of conjugation can be lost when we go to multiple rounds of error correction. 
    
    Fortunately this can be overcome by injecting `just enough' randomness -- the solution might be called `logical twirling' of the error channel (instead of twirling at the physical level). Logical twirling simply means twirling over the logical Pauli operators and decohere the Pauli components that corresponding to different logical operators. The resultant effective channel will be logically incoherent and thus the errors will accumulate at a more favourable rate in terms of logical fidelity. For one round of quantum error correction, applying logical twirling will not change the logical fidelity just like twirling a noise channel will not change its fidelity. Hence, applying logical fidelity on top of conjugation can maintain the fidelity improvement brought by conjugation in each round of error correction while preventing the logical errors from rapid accumulation as we go to multiple rounds. 
    
    The Pauli components of a given noise channel can be partitioned into sets that correspond to different logical operators with different measured syndromes after quantum error correction. As discussed in Section~\ref{sect:mech_conj}, the coherence between the components that correspond to the same logical operator and the same syndrome can be used by conjugation to improve the logical fidelity of the channel in one round of quantum error correction (via destructive interference between the logical error components), while in this section we see that the coherence between different logical operators can be removed by logical twirling to fight the accumulation of logical errors in multiple rounds of quantum error correction.
    
    In Appendix~\ref{sect:multi_same_coher}, using global $Z$ rotation as an example, we have demonstrated that using conjugation alone, the advantages of conjugation over physical twirling will diminish as we go to more rounds of quantum error correction and we have also shown how this is overcome by using logical twirling.

    \subsection{Multiple Rounds of Noise Tailoring}
    Instead of applying both noise tailoring and error correction at each time step, we can apply just noise tailoring in each time step and only do one round of error correction at the very end. 
    
    The matrix elements for the effective noise channel with $K$ rounds of twirling are:
    \begin{align}
    &\quad (\overline{\mathcal{N}}_{TK})_{G, G'} \nonumber\\
    &= \frac{1}{|\vec{\mathbb{W}}|}\sum_{\vec{W} \in \vec{\mathbb{W}}}\pbra{\Pi_{\vec{0}}\overline{G}} \mathcal{R} \left(\prod_{k=1}^{K}{\supop{W}}_k\right) \left(\prod_{k=K}^{1} \mathcal{N} {\supop{W}}_k\right)\pket{\Pi_{\vec{0}}\overline{G}'}
    \end{align}
    Here we divide the noise process into $K$ steps and apply a random Pauli gate $W_k$ at the beginning of each step. At the end, we undo all these random Pauli gate by applying the their inverse $\prod_{k=1}^K W_k$ and then perform quantum error correction. We denote the set of $K$ Pauli gate chosen using a vector $\vec{W}$. Similar to our arguments in Section~\ref{sect:twirling}, multiple rounds of twirling correspond to the average of all the Pauli conjugation schemes, thus one of the Pauli conjugations will be optimal and outperforms twirling.
    
    As detailed in Appendix~\ref{sect:multi_reduction}, if we want to find the equivalent conjugations to reduce the search space for multi-round conjugation, we can use similar arguments about the structure of the noise (Section~\ref{sect:noise_sym}) and the symmetries in both the noise and the code (Section~\ref{sect:perm_sym}), while the arguments about interaction with the code space to remove stabilisers and logical operators (Section~\ref{sect:remove_stb}) can only be applied to the outer-most round of conjugation.
    
    The search space of possible conjugations grows exponentially with the number of rounds while the number of symmetries that we can utilise is less than the one-round case (since we cannot remove all the stabilisers and logical operators from the twirling generating set). Hence, iterating over the whole search space might not be practical for a large number of rounds. However, we can still sample different conjugation schemes in our reduced search space to find a better scheme than doing nothing or even twirling, though such a scheme might not be optimal. 
    
    \section{Conclusion}\label{sect:conclusion}
    In this Article, we have shown that when doing one round of quantum error correction on a coherent noise channel, part of its coherence can actually be used to improve its logical fidelity using Pauli conjugation, which outperforms twirling. To search for the optimal Pauli conjugation under a given noise model using a given quantum error correction code, we use the properties and the symmetries of the noise and the code to identify the equivalent conjugations to reduce our search space. We applied our techniques to the Steane code, the Shor code and the surface code under a global $Z$ rotation noise, reducing the $4^n$ possibilities of Pauli conjugation to $2$, $4$ and $6$ equivalent classes respectively for those three codes. Iterated over these different classes of conjugations, we managed to find the optimal conjugations for each code, which resulted in higher logical fidelities than the twirled and original noise channel. We have shown via simulation that the advantages of the optimal Pauli conjugation schemes remain with gate errors present. Conjugation can also lead to higher concatenated thresholds than the twirled threshold. The conjugated threshold showed improvements over the original thresholds by $40\%$, $160\%$ and $110\%$ for the three codes we considered under the coherent $Z$ noise. We showed that by using logical twirling to remove the `harmful' coherence within the error channel, we can extend the advantages of Pauli conjugation to multiple rounds of error correction. We also briefly discussed how to extend our arguments into multiple rounds of Pauli conjugation.
    
    Compared to twirling, Pauli conjugations does not require the implementation of a random circuit, and the weights of the gates that we need to implement can be on average much smaller than twirling as shown by our examples. Being a deterministic scheme, it can be implemented in hardware systems in which modifying the circuit at each run is hard. It can also be used in quantum communication to combat coherent noise in the communication channel without needing to transmit the extra random bit needed by twirling. Single-qubit Pauli gates are usually the gates with the highest fidelity, combining with the fact that the Pauli conjugation gates we need to implement can be low-weight, it should be resilient to gate errors, as shown by our simulation. Hence, Pauli conjugation can be a practical way forward to mitigate errors in real experiments.
    
    The way we reduce the Pauli conjugation search space is highly dependent on the code we use and the noise model we have. Though our techniques work for the simple examples that we have considered, searching over all possible Pauli conjugations may not be feasible when the size of our system increase, when there are very few symmetries in the noise or when we are considering multiple rounds of Pauli conjugation. Hence, we might want to find a way to construct the optimal conjugation based on the mechanism of conjugation in Section~\ref{sect:mech_conj}, or at least find a better searching strategy than random sampling. Furthermore, we may not know the full noise model in practice. As discussed in Section~\ref{sect:mech_conj}, conjugation will only act on the coherent components in the channel, thus to find the optimal (or close-to optimal) conjugation gate we only need the information about the dominant coherent component in the channel without needing any information about the incoherent parts or the other small coherent components. In the worst case scenario, we can still sample over the different Pauli conjugations based on any limited information we have to find a scheme with better performance than the original noise channel instead of finding the optimal one. 
    
    The above ideas can be tested by applying Pauli conjugation to more general error channels beyond the global $Z$ rotation. An example will be the general local $Z$ noise channel considered in~\cite{huangPerformanceQuantumError2019} or some non-biased noise models like those considered in~\cite{gutierrezErrorsPseudothresholdsIncoherent2016}. To see if the conjugation technique is valuable in fault-tolerant computation, it will also be interesting to see how Pauli conjugation will perform against gate-level coherent noise and whether it can improve the surface code threshold (instead of the concatenated threshold) given a realistic noise model.
    
    There are several degrees of freedom we can add to further optimise our noise tailoring schemes. Firstly, throughout this Article we have been focusing on conjugation using Pauli gates, it will be interesting to extend our technique to Clifford gates or even general unitaries. We can also look into the case where we allow Clifford correction~\cite{chamberlandHardDecodingAlgorithm2017}. We definitely did not exhaust all the ways to reduce the Pauli conjugation search space. For example, we have only been focusing on the permutation symmetry of code and noise, which at best can only prove the operators with the same weight are equivalent. A next step could be including other Clifford symmetries like CZ gates, etc.
    
    Our conjugation scheme, especially the multi-round variant, in a way can be viewed as bang-bang dynamical decoupling tailored to a given quantum error correction code.  Attempts has been made before to study the effect of dynamical decoupling within the context of quantum error correction~\cite{ngCombiningDynamicalDecoupling2011}, though without explicitly considering the code structure like we did in this Article. It will be a fruitful area to adapt more schemes in the established literature of dynamical decoupling~\cite{suterColloquiumProtectingQuantum2016} into the context of quantum error correction taking into account the code structure. We may get a fuller understanding about how to search for better multi-round conjugation scheme from the way we optimise dynamical decoupling using average Hamiltonian arguments~\cite{violaDynamicalDecouplingOpen1999} and group theoretic arguments~\cite{zanardiSymmetrizingEvolutions1999}. Ideas like non-equidistant pulses~\cite{uhrigKeepingQuantumBit2007}, robust decoupling sequences and higher-order decoupling~\cite{suterColloquiumProtectingQuantum2016} can also be extended into multi-round conjugation. 
    
    Besides applications in quantum error correction for memory, the conjugation technique can also be extended into other fields like quantum metrology and quantum simulation. For quantum metrology with error correction~\cite{kesslerQuantumErrorCorrection2014,durImprovedQuantumMetrology2014,zhouAchievingHeisenbergLimit2018}, we hope to find conjugation schemes that can tailor the noise into a form that is less damaging to the code and/or tailor the signal into a form that the code is more sensitive towards. When applied to symmetry verification in quantum simulation~\cite{bonet-monroigLowcostErrorMitigation2018a, mcardleErrorMitigatedDigitalQuantum2019, mccleanDecodingQuantumErrors2019}, conjugation may enable more noise to be detected via transformation of the previously undetected noise components. In the above applications, it is likely that we need to develop more complex conjugation schemes beyond one-round Pauli conjugation.
    
    \section*{Acknowledgements}
    The authors are grateful to Markus M\"uller for helpful and encouraging conversations. ZC thanks Y. Li for valuable discussions and E. Huang for providing the equations for the noise map used in the calculation of the concatenated thresholds. 
    
    ZC acknowledges support from Quantum Motion Technologies Ltd. 
    
    SCB acknowledges support from ESPRC grant EP/M013243/1 (the NQIT Quantum Hub). 
    
    SCB and XX are supported by the Office of the Director of National Intelligence (ODNI), Intelligence Advanced Research Projects Activity (IARPA), via the U.S.
    Army Research Office Grant No. W911NF-16-1-0070.
    The views and conclusions contained herein are those of
    the authors and should not be interpreted as necessarily representing the official policies or endorsements, either expressed or implied, of the ODNI, IARPA, or the
    U.S. Government. The U.S. Government is authorized to
    reproduce and distribute reprints for Governmental purposes notwithstanding any copyright annotation thereon.
    Any opinions, findings, and conclusions or recommendations expressed in this material are those of the author(s)
    and do not necessarily reflect the view of the U.S. Army
    Research Office.
    
    \section*{Competing Interests}
    The authors declare no conflict of interest.
    
    \section*{Data Availability}
    The data sets generated during the current study are available from the corresponding author on reasonable request.
    
    \section*{Contributions}
    ZC performed the theoretical analysis and wrote the manuscript with input from SCB. XX performed the numerical simulation for the circuits with gate errors, ZC performed the other numerical simulations. 
    
    \appendix
    \section{Construction of $\widetilde{\mathbb{E}} $}\label{sect:construct_E}
    Recall that the error generators $\widetilde{\mathbb{E}}$ are just all the remaining generators needed to complement the stabiliser generators and the logical generators for generating the full Pauli gate set. The requirements of $\widetilde{\mathbb{E}}$ are 
    \begin{enumerate}
        \item All of its elements are independent.
        \item The size of $\widetilde{\mathbb{E}}$ is $|\widetilde{\mathbb{E}}| = |\widetilde{\mathbb{S}}|$.
        \item The full set of elements that can be generated by $\widetilde{\mathbb{E}}$ does not contain any elements in $\mathbb{S}$ or $\overline{\mathbb{G}}$, otherwise we can replace the generators in $\widetilde{\mathbb{E}}$ with the elements in $\mathbb{S}$ or $\overline{\mathbb{G}}$.
    \end{enumerate}
    The full Pauli set $\mathbb{G}$ can be generated using all the single physical qubit $X$ and $Z$ gate, after removing the elements that are dependent on each other through composition with stabiliser generators and/or logical generators, we will be left with the generating set $\widetilde{\mathbb{E}}$. Hence, we can always find a $\widetilde{\mathbb{E}}$ that consist of only single qubit $X$ or $Z$ operators. Here we will show how do we construct it.
    
    Any practical stabiliser error correction code will be able to detect and correct all single qubit $X$ and $Z$ errors, hence all of these single-qubit errors will violate different subsets of stabiliser checks. The way we construct $\widetilde{\mathbb{E}}$ is:
    \begin{enumerate}
        \item Find all single-qubit $X$ and $Z$ errors that violate only one stabiliser check and add them to $\widetilde{\mathbb{E}}$. We will denote the set of stabiliser checks that they violate as $\mathbb{S}_E$.
        \item Starting with $n = 2$, search in the checked physical qubits of the stabiliser checks in $\mathbb{S}_E$, there will be $X$ or $Z$ errors on these qubits that fail $n$ stabiliser checks, with one and only one of the failed stabiliser checks \emph{not} in $\mathbb{S}_E$. For each of such error we found, we will add it into $\widetilde{\mathbb{E}}$ and add the one additional violated stabiliser check into $\mathbb{S}_E$. Note that for each new element added into $\mathbb{S}_E$, we will have more physical qubits to check.
        \item Repeat step 2 with $n$ increasing by 1 in each iteration until $\mathbb{S}_E = \widetilde{\mathbb{S}}$, i.e. until $\mathbb{S}_E$ contains all the stabiliser checks (or equivalently until $|\widetilde{\mathbb{E}}| = |\widetilde{\mathbb{S}}|$).
    \end{enumerate}
    In the case of topological code with boundaries, the above scheme is just starting by adding the stabiliser checks at the boundary into $\mathbb{S}_E$ and slowly progressing inwards, adding the inner stabiliser checks into $\mathbb{S}_E$ until all stabiliser checks are within $\mathbb{S}_E$.
    
    In this way of construction, there is no way to find any composition of elements in $\widetilde{\mathbb{E}}$ such that there are no stabiliser checks fail, hence there is no way to compose stabilisers or logical operators out of these elements. 
    
    \section{Construction of $\mathbb{W}$}\label{sect:W_construction}
    With the twirling generators $\widetilde{\mathbb{W}}$ obtained in Section~\ref{sect:noise_sym}, we can now generate the full set of twirling gate $\mathbb{W}$. The elements in the twirling set $\mathbb{W}$ will correspond to the error operators that are detectable by our quantum error correction code and will all have different syndromes. For the purpose of twirling, we would want to replace these operators with the lowest weight error operators that produce the same syndrome (i.e. equivalent up to composition with stabilisers and logical operators), since operators with lower weight will be easier to implement with fewer errors induced. These are usually just the recovery operators of the given syndromes, in such case we can just get them from the decoder. For a distance-$d$ code, by definition all errors with weight $d-1$ or lower produce non-trivial error syndromes, and all errors with weight $\lfloor\frac{d - 1}{2}\rfloor$ or below will have different syndromes (thus correctable). Hence, for all $W \in \mathbb{W}$ with weight $\lfloor\frac{d - 1}{2}\rfloor$ or below, they are already the lowest weight operators that can produce the given syndrome, while for the others in $\mathbb{W}$ it may be possible to find a lower weight equivalence (not guaranteed to find since some correctable errors can be of higher weight than $\lfloor\frac{d - 1}{2}\rfloor$, e.g. a surface code with very long $Z$ boundaries and very short $X$ boundaries).
    
    \section{Twirling Generators Reduction}\label{sect:generator_reduce}
    Using $\supsupop{\quad}$ to denote `super-super-operators':
    \begin{align*}
    \supsupop{A} (\supop{C}) = \supop{A}\supop{C}\supop{A^\dagger} 
    \end{align*}
    we can rewrite that twirling process as:
    \begin{align*}
    \mathcal{T}(\mathcal{N}) & = \frac{1}{\abs{\mathbb{W}}}\sum_{W \in \mathbb{W}} \supsupop{W} (\mathcal{N})\\
    & = \prod_{W \in \widetilde{\mathbb{W}}} \frac{I + \supsupop{W}}{2} \mathcal{N}
    \end{align*}
    \begin{align}
    \overline{\mathcal{N}}_T &= \mathcal{R}\mathcal{T}(\mathcal{N})\nonumber\\
    & = \mathcal{R} \prod_{W \in \widetilde{\mathbb{W}}} \frac{I + \supsupop{W}}{2} \mathcal{N} \label{eqn:twirled_channel}
    \end{align}
    Here we have implicitly assumed that $\supsupop{W}$ only acts on $\mathcal{N}$: $\supsupop{W} \mathcal{N} = \supsupop{W}(\mathcal{N})$.
    
    Hence, the matrix elements for the twirled logical channel are
    \begin{align*}
    \overline{\mathcal{N}}_{T, GG'} & = \frac{1}{2^{|\widetilde{\mathbb{G}}|}}\pbra{\overline{G}\Pi_0} \mathcal{R} \prod_{W \in \widetilde{\mathbb{G}}} \left(I + \supsupop{W}\right)\mathcal{N} \pket{\overline{G}'\Pi_0} 
    \end{align*}
    All Pauli super-operators commute, thus all Pauli super-super-operators also commute. Hence, we can arrange the order of the twirling generators in $\prod_{W \in \widetilde{\mathbb{G}}} \left(I + \supsupop{W}\right)$ in any way we want. We will arrange it in the following way
    \begin{align*}
    \frac{1}{2^{|\widetilde{\mathbb{G}}|}}\prod_{W \in \widetilde{\mathbb{G}}} \left(I + \supsupop{W}\right) &= \frac{1}{2^{|\widetilde{\mathbb{G}}|}}\prod_{S \in \widetilde{\mathbb{S}}} \left(I + \supsupop{S}\right) \prod_{\overline{G} \in \widetilde{\overline{\mathbb{G}}}} \left(I + \supsupop{\overline{G}}\right) \\
    &\quad \times \prod_{E_n \in \widetilde{\mathbb{E}}_n} \left(I + {\supsupop{E}}_n\right)  \prod_{E_c \in \widetilde{\mathbb{E}}_c} \left(I + {\supsupop{E}}_c\right)
    \end{align*}
    where $\widetilde{\mathbb{E}}_c$ is a subset of $\widetilde{\mathbb{E}}$ that acts trivially on noise $\mathcal{N}$ when used for twirling as discussed in Section~\ref{sect:noise_sym}. 
    
    Thus $\frac{1}{2^{|\widetilde{\mathbb{E}}_c|}} \prod_{E_c \in \widetilde{\mathbb{E}}_c} \left(I + {\supsupop{E}}_c\right)$ will act trivially on $\mathcal{N}$ and can be absorbed by $\mathcal{N}$ when we put them closest to $\mathcal{N}$. 
    
    On the other hand, the twirling of $\widetilde{\mathbb{S}}$ and $\widetilde{\overline{\mathbb{G}}}$ will act trivially on the error correction code and the logical states (see Section~\ref{sect:remove_stb}), hence we can put them nearest to the logical states to remove them.
    
    In Section~\ref{sect:perm_sym}, the equivalence of the twirling gates is obtained via interaction with both the noise elements and the logical states. To allow a generator to interact with both the noise elements and the logical states, we need to permute the symmetry operator of the noise elements or the logical states through the other generators, which will modify them. Hence instead of proving equivalence of generators, we expand out the product of generators to obtain a linear combination of the elements in the twirling set, and prove their equivalence instead.
    
    \section{Derivation of Eqn~(\ref{eqn:equiv_symmetry})}\label{sect:perm_sym_deriv}
    If all the code state basis $\Pi_{\vec{0}}\overline{G}$ and the physical noise channel $\mathcal{N}$ are invariant under the Clifford transformation $U$:
    \begin{align*}
    \left[U, \Pi_{\vec{0}}\overline{G}\right] &= 0 \quad \forall G \in \mathbb{G}\\
    \left[\supop{U}, \mathcal{N}\right] &= 0
    \end{align*}
    then the recovery channel $\mathcal{R}$ will also be invariant under the same transformation since it is completely based on the code and the error channel: $\left[\supop{U}, \mathcal{R}\right] = 0$. 
    
    Hence, we have:
    \begin{align*}
    \left[U, \Pi_{\vec{0}}\overline{G}\right] &= 0 \quad \forall G \in \mathbb{G}\\
    \left[\supop{U}, \mathcal{N}\right] &= 0,\ \left[\supop{U}, \mathcal{R}\right] = 0 \Rightarrow \left[\supop{U}, \mathcal{R}\mathcal{N}\right] = 0
    \end{align*}
    Thus
    \begin{align*}
    U\Pi_{\vec{0}}\overline{G}U^\dagger &= \Pi_{\vec{0}}\overline{G} \quad \Rightarrow\quad \supop{U} \pket{\Pi_{\vec{0}}\overline{G}} = \pket{\Pi_{\vec{0}}\overline{G}} \quad \forall G \in \mathbb{G}\\
    \supop{U}\mathcal{R}\mathcal{N}\supop{U^\dagger} &= \mathcal{R}\mathcal{N}
    \end{align*}
    Since $\supop{W}$ and $\mathcal{R}$ are both Pauli channel, they commutes, hence
    \begin{align*}
    \overline{\mathcal{N}}(W)_{G, G'} & = \pbra{\Pi_{\vec{0}}\overline{G}} \mathcal{R}\supop{W} \mathcal{N} \supop{W} \pket{\Pi_{\vec{0}}\overline{G}'}\\
    & = \pbra{\Pi_{\vec{0}}\overline{G}} \supop{W} \mathcal{R} \mathcal{N} \supop{W} \pket{\Pi_{\vec{0}}\overline{G}'}\\
    & = \pbra{\Pi_{\vec{0}}\overline{G}}\supop{U^\dagger}\supop{W} \supop{U}\mathcal{R} \mathcal{N} \supop{U^\dagger}\supop{W} \supop{U}\pket{\Pi_{\vec{0}}\overline{G}'}
    \end{align*}
    Since $U$ is Clifford, $U^\dagger W U$ is Pauli, hence $\supop{U^\dagger W U}$ commute with $\mathcal{R}$:
    \begin{align*}
    \overline{\mathcal{N}}(W)_{G, G'} & = \pbra{\Pi_{\vec{0}}\overline{G}}\mathcal{R} \supop{U^\dagger W U} \mathcal{N} \supop{U^\dagger W U}\pket{\Pi_{\vec{0}}\overline{G}'}\\
    & = \overline{\mathcal{N}}(U^\dagger W U)_{G, G'}
    \end{align*}
    
    \section{Shape of Fidelity Curve for Global $Z$ Rotation}\label{sect:shape_fidelity}
    \subsection{Rotational symmetry of the fidelity curve}
    For the noise model in (\ref{eqn:noise_model}) and for all the codes we considered which have global logical $Z$ gates, we have:
    \begin{align*}
    N(\frac{\pi}{2}) = \prod_{j = 1}^J (-iZ_j) \equiv \overline{Z}
    \end{align*}
    For the worst case fidelity $Q$ for such pure $Z$ noise, we will start and measured in the logical $\ket{+_L}$ eigenstate:
    \begin{align*}
    Q(\theta) &= \abs{\bra{+_L}N(\theta) \ket{+_L}}^2 \\
    &= \abs{\left(\bra{+_L}N(\theta) \ket{+_L}\right)^*}^2\\
    &= \abs{\bra{+_L}N(-\theta) \ket{+_L}}^2
    \end{align*}
    \begin{align*}
    Q(\frac{\pi}{2}-\theta) &= \abs{\bra{+_L}N(\frac{\pi}{2}-\theta) \ket{+_L}}^2 \\
    &= \abs{\bra{+_L}N(\frac{\pi}{2})N(-\theta) \ket{+_L}}^2 \\
    &= \abs{\bra{-_L}N(-\theta) \ket{+_L}}^2 \\
    &= 1 -  \abs{\bra{+_L}N(-\theta) \ket{+_L}}^2\\
    &= 1 -  Q(\theta)\\
    \end{align*}
    For fidelity of one qubit, we have:
    \begin{align*}
    F(\theta) = \frac{2}{3} Q(\theta) + \frac{1}{3}
    \end{align*}
    Hence, we have:
    \begin{align*}
    F(\frac{\pi}{2}-\theta) & = \frac{2}{3} \left(1 -  Q(\theta)\right) + \frac{1}{3}\\
    & = \frac{4}{3} -  \frac{2}{3}Q(\theta) - \frac{1}{3}\\
    & = \frac{4}{3} - F(\theta)
    \end{align*}
    Hence, the logical fidelity curve $F(\theta)$ is rotationally symmetry about a point at $\theta = \frac{\pi}{4}$
    
    \subsection{Fidelity at $\theta = \frac{\pi}{4}$}
    For the noise model in (\ref{eqn:noise_model}), at $\theta = \frac{\pi}{4}$ we have:
    \begin{align}\label{eqn:half_Z}
    N(\frac{\pi}{4}) = \frac{1}{\sqrt{2^J}}\prod_{j = 1}^J \left( I - iZ_j \right)
    \end{align}
    For an operator $U$ consist of tensor product of single-qubit $Z$, we will write the set of qubit index that we apply $Z$ gate on as $\vec{U}$:
    \begin{align*}
    U = \prod_{i \in \vec{U}} Z_i
    \end{align*}
    Hence, the terms in the expansion of (\ref{eqn:half_Z}) will be $ (-i)^{\abs{\vec{U}}} \prod_{i \in \vec{U}}Z_i = (-i)^{\abs{\vec{U}}} U$. In the case of measuring the zero syndrome, $N(\frac{\pi}{4})$ will collapse into a superposition of stabilisers and $Z$ logical operators. For each stabiliser term $(-i)^{\abs{\vec{S}}} S$ in the expansion, there will be a corresponding logical $Z$ operator term differed by apply $Z$ to all qubits: $(-i)^{J - \abs{\vec{S}}} S \left(\prod_j Z_j\right) = (-i)^{J - \abs{\vec{S}}} S \overline{Z}$, hence the terms in the expansion correspond to zero syndrome is:
    \begin{align*}
    &\quad \sum_S (-i)^{\abs{\vec{S}}} S + \overline{Z} \left(\sum_{S} (-i)^{J - \abs{\vec{S}}} S\right)\\
    & = \sum_{S} (-i)^{\abs{\vec{S}}} S \left[ I + (-i)^{J - 2\abs{\vec{S}}} \overline{Z}\right]\\
    & \equiv \sum_{S} \left[ \overline{I} + (-i)^{J - 2\abs{\vec{S}}} \overline{Z}\right]
    \end{align*}
    If all stabilisers have even weights, i.e. $|\vec{S}| = 2n$, then $2|\vec{S}| = 4n$, hence $(-i)^{J - 2\abs{\vec{S}}} = (-i)^{J} (-i)^{-2\abs{\vec{S}}} = (-i)^{J}$. 
    
    Thus if all the stabilisers have even weights, and the logical $Z$ operator consist of applying $Z$ to all the qubits, then the terms in the expansion of (\ref{eqn:half_Z}) is:
    \begin{align*}
    \abs{\mathbb{S}} \left[ \overline{I} + (-i)^{J} \overline{Z}\right] 
    \end{align*}
    and similarly for other syndromes. Note that terms of other syndromes will also result in the same amplitude $\abs{\mathbb{S}}$. Hence, we have the same probability of collapse into any syndrome. For odd number of qubits $J$, we then have a logical $Z$ rotation of the angle $\frac{\pi}{2}$ (or $-\frac{\pi}{2}$), which means a worst case fidelity of $\frac{1}{2}$ and an average fidelity of $\frac{2}{3} * \frac{1}{2} + \frac{1}{3} = \frac{2}{3}$. 
    
    \section{Equivalent Conjugation Classes for Codes in Global $Z$ Rotation}\label{sect:other_code_details}
    Recalled the arguments in Section~\ref{sect:Z_noise_sim}. We can make the following simplification based on the codes and the noise model we consider.
    \begin{itemize}
        \item Our noise model is global, thus have all possible qubit permutation symmetry. Along with the fact that our codes have one logical qubit and global Pauli gates mean that we only need to consider the permutation symmetry of the stabilisers when we try to reduce the twirling set.
        \item The noise model is pure $Z$ noise, thus all $Z$ twirling generators can be removed. All $Z$ stabiliser checks will also have trivial effects (besides the five-qubit code which does not have pure $Z$ checks), thus we only need to consider the permutation symmetry of the $X$ stabilisers.
    \end{itemize}
    
    \subsection{Five-qubit code}
    In five-qubit code, the generators are
    \begin{itemize}
        \item Stabiliser generators $\widetilde{\mathbb{S}}$: $XZZXI$ and three of its cyclic permutations $IXZZX$, $XIXZZ$, $ZXIXZ$
        \item Logical generators $\widetilde{\overline{\mathbb{G}}}$: $X$ or $Z$ on all qubits.
    \end{itemize}
    Following Section~\ref{sect:remove_stb}, we can construct the twirling generators:
    \begin{align*}
    \widetilde{\mathbb{W}} = \widetilde{\mathbb{E}} = \{X_1, X_2, Z_3, Z_5\}
    \end{align*}
    As mentioned above, the $Z$ twirling generators can be safely removed since we have pure $Z$ noise:
    \begin{align*}
    \widetilde{\mathbb{W}} = \{X_1, X_2\}
    \end{align*}
    which generates the twirling set:
    \begin{align*}
    \mathbb{W}= \{I, X_1, X_2, Z_4\}
    \end{align*}
    Here we have transform the error operators $X_1X_2$ to its lowest weight equivalence with the same error syndromes $Z_4$. Conjugating the noise with $Z_4$ has trivial effect since we have pure $Z$ noise.
    
    The five-qubit code has cyclic permutation symmetry (there are also additional symmetries that we do not need to use here~\cite{huangPerformanceQuantumError2019}). As discussed in Section~\ref{sect:noise_sym}, using these symmetry transformation, we can easily prove that conjugating the noise with $X_1$ is equivalent to $X_2$ since $X_2 = U^\dagger X_1 U$ where $U$ is one of the qubit cyclic permutation operator.
    
    Hence, there are two equivalent class of twirling gate, one is equivalent to $I$, while the other is equivalent to $X_1$ (or any single-qubit $X$ gate by cyclic permutation). 
    
    \subsection{Nine-qubit Shor code}
    \subsubsection{With Local Z checks}
    As shown in Figure~\ref{fig:nine_shor}, in nine-qubit Shor Code, the generators are
    \begin{itemize}
        \item Stabiliser generators $\widetilde{\mathbb{S}}$: $\{Z_iZ_{i+1}\ |\ i \in \{1, 2, 4, 5, 7, 8\}\}$ and  $\{\prod_{j = 0}^{5} X_{i+j}\ |\ i \in \{1, 4\}\}$
        \item Logical generators $\widetilde{\overline{\mathbb{G}}}$: $X$ or $Z$ on all qubits. 
    \end{itemize}
    Following Section~\ref{sect:remove_stb}, we can construct our twirling generators:
    \begin{align*}
    \widetilde{\mathbb{W}} = \widetilde{\mathbb{E}} = \{X_1, X_3, X_4, X_6, X_7, X_9, Z_1, Z_7\}
    \end{align*}
    As mentioned above, the $Z$ twirling generators can be safely removed since we have pure $Z$ noise:
    \begin{align*}
    \widetilde{\mathbb{W}} = \{X_1, X_3, X_4, X_6, X_7, X_9\}
    \end{align*}
    When looking at the $Z$ stabiliser checks, which will produce the syndromes for these $X$ error operators, we realise they are divided into 3 non-overlapping set (no shared checked qubits), which are individual rows in Figure~\ref{fig:nine_shor}. All the error syndromes within each row can be produced by single-qubit $X$ errors within that row. The Z-check syndrome of different rows are independent of each other since they do not share any qubits. Hence, to produce all possible syndromes using error operators with the lowest weight, we will have zero or one single-qubit X errors in each row. This set of error operators will be the full twirling set $W$ that is generated.
    
    Permutation symmetries that exist in the 9-qubit Shor code will be any permutation of the elements within each row and any permutation between the rows shown in Figure~\ref{fig:nine_shor}. In such a case, all the operators with the same weight in our twirling set can be shown to be equivalent, leaving us with the following four equivalent classes of twirling operators.
    \begin{itemize}
        \item Identity: $I$
        \item Single qubit flip: $X_1$
        \item Two-qubit flip (in different row): $X_1X_4$
        \item Three-qubit flip (in different row): $X_1X_4X_7$
    \end{itemize}
    \subsubsection{With Local X checks}
    It is still a nine-qubit Shor code, with a swap between the $X$ and $Z$ stabilisers. 
    
    Following Section~\ref{sect:remove_stb}, we can construct our twirling generators:
    \begin{align*}
    \widetilde{\mathbb{W}} = \widetilde{\mathbb{E}} = \{Z_1, Z_3, Z_4, Z_6, Z_7, Z_9, X_1, X_7\}
    \end{align*}
    As mentioned above, the $Z$ twirling generators can be safely removed since we have pure $Z$ noise:
    \begin{align*}
    \widetilde{\mathbb{W}} = \{X_1, X_7\}
    \end{align*}
    which generates the twirling set:
    \begin{align*}
    \mathbb{W}= \{I, X_1, X_4, X_7\}
    \end{align*}
    Here we have transform the error operators $X_1X_7$ to its lowest weight equivalence with the same error syndromes $X_4$.
    
    We have the same code symmetry as the other Shor code, which allows us to prove the equivalence of conjugation with $X_1$, $X_4$ and $X_7$. 
    
    Hence, there are two equivalent class of twirling gate, one is equivalent to $I$, while the other is equivalent to $X_1$ (or any single-qubit $X$ gate). 
    
    \subsection{Distance-3 surface code}
    The stabiliser generators of the distance-3 surface code is shown in Figure~\ref{fig:surface_code}.
    
    Following Section~\ref{sect:remove_stb}, we can construct our twirling generators:
    \begin{align*}
    \widetilde{\mathbb{W}} = \widetilde{\mathbb{E}} = \{X_1, X_3, X_7, X_9, Z_1, Z_3, Z_7, Z_9\}
    \end{align*}
    As mentioned above, the $Z$ twirling generators can be safely removed since we have pure $Z$ noise:
    \begin{align*}
    \widetilde{\mathbb{W}} = \{X_1, X_3, X_7, X_9\}
    \end{align*}
    which generates the twirling set:
    \begin{itemize}
        \item Weight-1: $X_1$, $X_2$, $X_3$, $X_5$, $X_7$, $X_8$, $X_9$
        \item Weight-2: $X_2X_7$, $X_1X_7$, $X_1X_8$, $X_1X_9$, $X_2X_9$, $X_3X_9$, $X_3X_8$, $X_2X_8$
    \end{itemize}
    The logical Pauli gates of the code are just applying the corresponding Pauli gates to all the physical qubits (because it is an odd distance surface code), hence we only need to look at the symmetry of its stabilisers as discussed in Section~\ref{sect:perm_sym}. We can group the operators that are equivalent due to the rotational symmetry of the code: ($X_1$, $X_9$), ($X_2$, $X_8$), ($X_3$, $X_7$), ($X_5$), ($X_2X_7$, $X_3X_8$), ($X_1X_7$, $X_3X_9$), ($X_1X_8$, $X_2X_9$), ($X_1X_9$), ($X_2X_8$).
    
    Since we have only pure $Z$ noise, we only need to look at the symmetry exists in the X stabilisers, leading to additional symmetry in the exchange between qubits ($1$, $2$) and between qubits ($8$, $9$). Applying on top of the rotational symmetry, we have the following classes of equivalent conjugations:
    \begin{itemize}
        \item $I$
        \item $X_1$, $X_2$, $X_8$, $X_9$
        \item $X_3$, $X_7$
        \item $X_5$
        \item $X_1X_7$, $X_3X_9$, $X_2X_7$, $X_3X_8$
        \item $X_1X_9$, $X_2X_8$, $X_1X_8$, $X_2X_9$
    \end{itemize}

    \section{Effective $Z$ Logical Channel Conditioned on Syndrome}\label{sect:sign_log_rot}
    When we expand the physical noise $e^{- i\theta \sum_j Z_j}$ into the sum of tensor products of $Z$, all the odd-weight term will form the imaginary part, while all even-weight terms will form the real part. Hence, when we flip the sign of $\theta$, which is equivalent to taking the complex conjugate of our noise channel, all the odd-weight terms (the imaginary part) will flip their signs while all the even-weight terms (the real part) will remain the same.
    
    By saying the operator is real (imaginary) here, we mean that the operator has real (imaginary) amplitude. Since we are only considering $Z$ noise, composing two $Z$ operators together will not lead to any extra phase factor $i$. Thus when we compose an imaginary $Z$ operator with a real $Z$ operator, we will get an imaginary $Z$ operator, while composing two imaginary or two real operators will give a real $Z$ operator.
    
    For all the codes we consider here, they have stabiliser generators that are all even-weight, which means that they are all real in the noise expansion. For our codes, we can find one of the logical operators $\overline{Z}$ is odd-weight, which corresponds to imaginary amplitude in the expansion, thus all $\overline{Z}$ are imaginary since they can be obtained by composing the imaginary $\overline{Z}$ with the real stabilisers. Hence, when we measured the $0$ syndrome, the noise is collapsed into a coherent superposition of $\overline{I}$ with real amplitude and $\overline{Z}$ with imaginary amplitude. When we flip the sign of the physical error angle $\theta$, it is the same as taking the complex conjugate, which will flip the sign of $\overline{Z}$ since its amplitude is pure imaginary. For other syndromes with error $E$, we still have one of $E\overline{Z}$ and $E\overline{I}$ being one real and the other being imaginary and hence similar argument follows.
    
    For codes with odd-weight stabiliser generators, we will have a complex amplitude (mix of real and imaginary) for $\supop{I}$. On the other hand, for codes with even-weight logical operators (even distance code), it will give real $\overline{Z}$ for real $\overline{I}$ and imaginary $\overline{Z}$ for imaginary $\overline{I}$. In both case, since the phase of $\overline{I}$ and $\overline{Z}$ are no longer guaranteed to be differed by $i$, the logical channel for a given syndrome can no longer be written as a logical $\overline{Z}$ rotation, but instead a combination of logical $\overline{Z}$ rotation and logical dephasing channel. This was observed by Huang \textit{et. al.}~\cite{huangPerformanceQuantumError2019} for even-distance repetition code and distance-4 surface code.
    
    \section{Pauli Conjugation for the Other 9-qubit Shor Code}\label{sect:other_shor}
\begin{figure}[htbp!]
        \centering
        \includegraphics[width=0.5\textwidth]{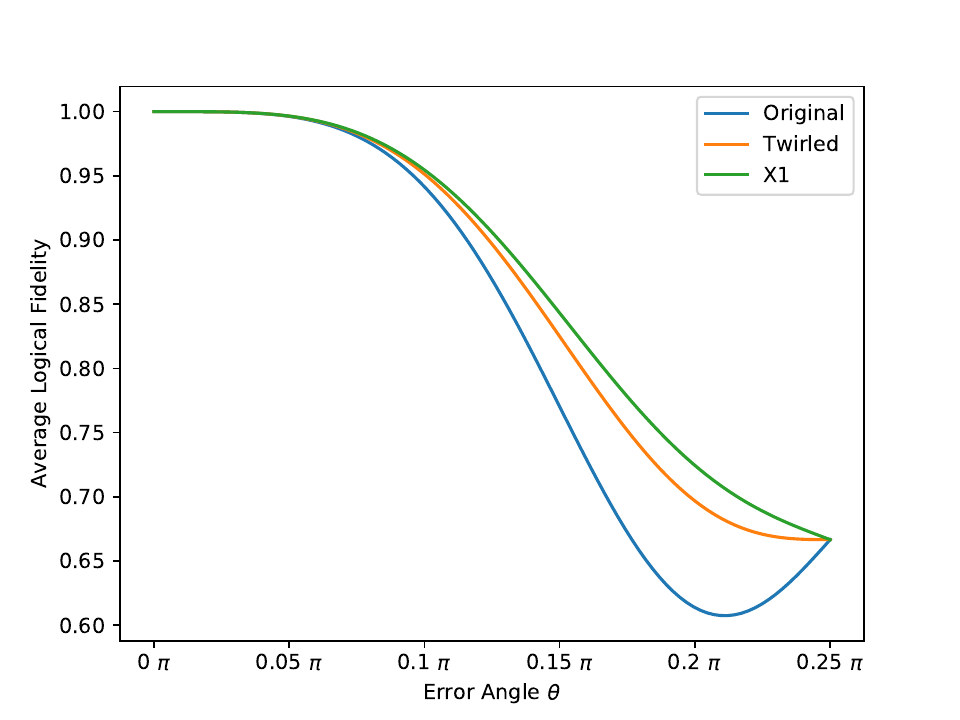}
        \caption{Logical fidelity of the nine-qubit shor code with local X checks under difference noise strength and noise tailoring schemes.}
        \label{fig:shor_code_fidelity_2}
    \end{figure}

\begin{figure}[htbp!]
        \centering
        \includegraphics[width=0.5\textwidth]{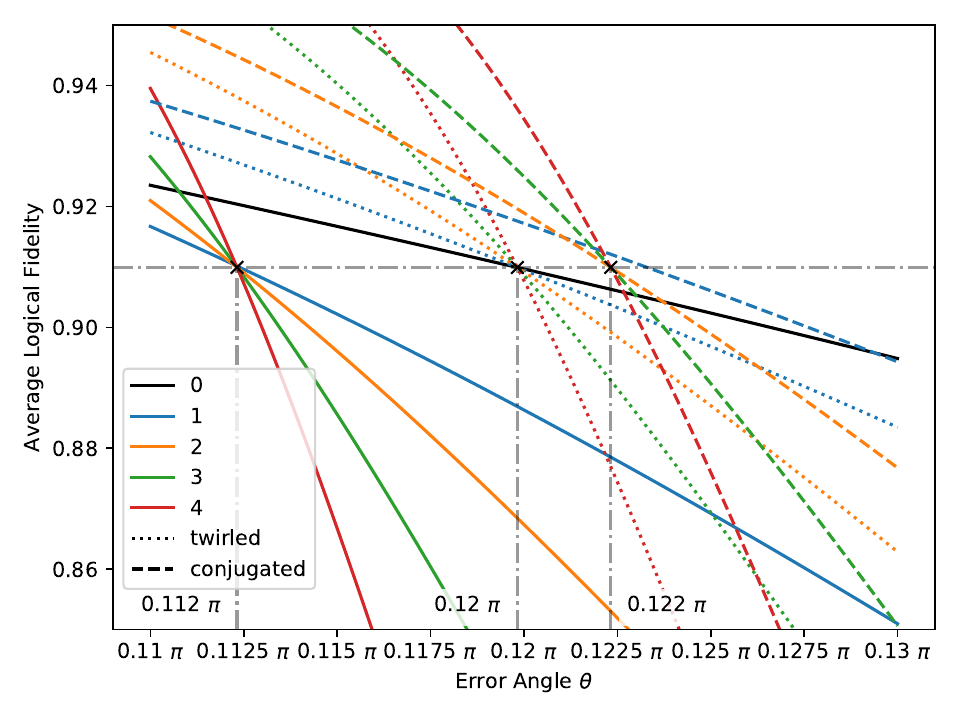}
        \caption{The concatenated threshold plot under global $Z$ rotation noise for the 9-qubit Shor code with local X checks. Different colours show different levels of concatenation while different line styles show applying different strategies like twirling or Pauli conjugation to the noise. The Pauli conjugations we used here are the optimal schemes that we found with zero gate error.}
        \label{fig:shor_threshold_plots}
    \end{figure}

    In the main text, we have only shown results for the 9-qubit Shor code local $Z$ checks. For the 9-qubit Shor code with local $X$ check, its average fidelity of different conjugation schemes and its concatenated threshold plots are shown in Figure~\ref{fig:shor_code_fidelity_2} and \ref{fig:shor_threshold_plots}.
    
\begin{figure}[htbp!]
        \centering
        \includegraphics[width=0.5\textwidth]{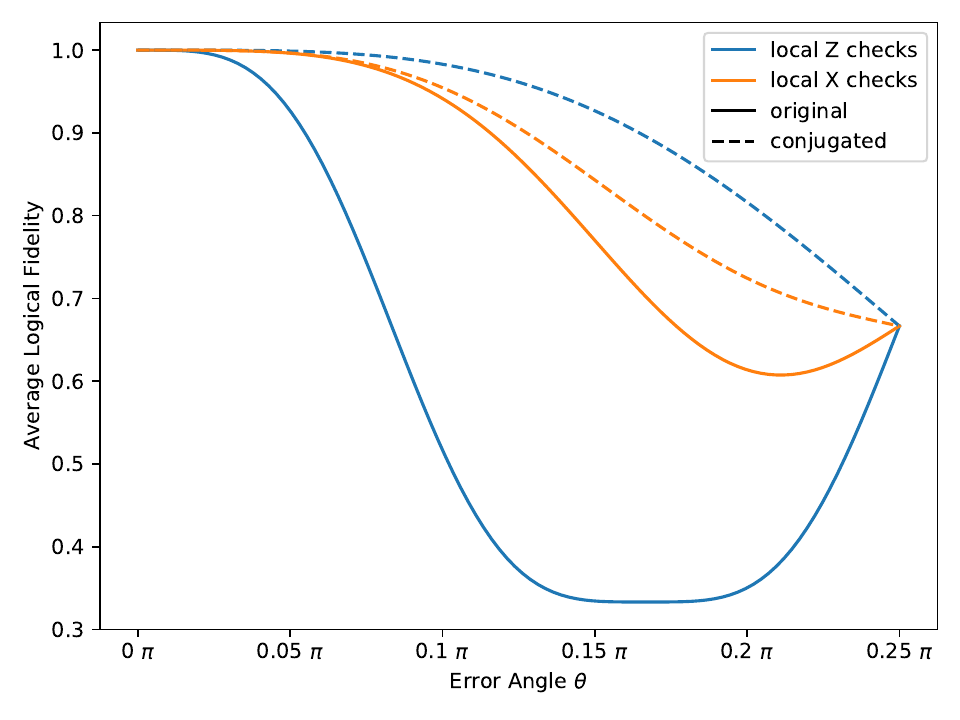}
        \caption{Comparison between Shor code with local X checks and local Z checks with Pauli conjugation and without. The Pauli conjugations we used here are the optimal schemes that we found with zero gate error.}
        \label{fig:shor_comparison}
    \end{figure}

    For the $Z$ noise we considered, the 9-qubit Shor code that has local $X$ checks will actually have better performance since it has more $X$ checks which are sensitive to $Z$ noise. This is shown by the large gap between the two original fidelity curve in Figure~\ref{fig:shor_comparison}. However, after using conjugation to tailor the noise to fit the code, the Shor code with local $Z$ checks receives a huge boost in fidelity such that it even exceeds the fidelity of the other Shor code with conjugation. This further exemplifies the power of Pauli conjugation when there is a misfit between the code and the noise.
    \section{Detailed Circuit for the Codes}\label{sect:encoding_circ}
\begin{figure}[htbp!]
        \centering
        \includegraphics[width=0.5\textwidth]{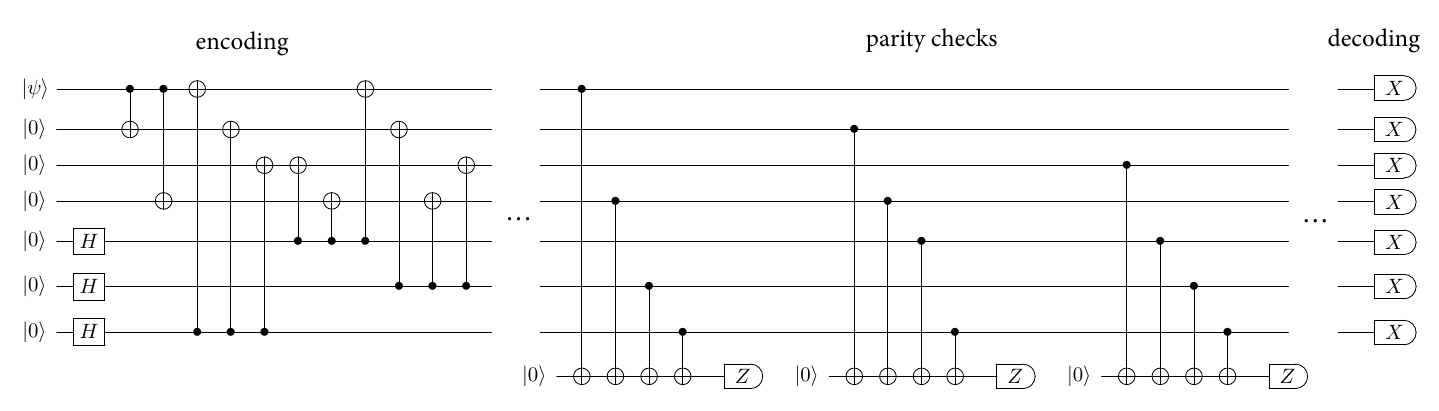}
        \caption{The encoding and the parity check circuit for the Steane code.}
        \label{fig:steane_encode}
    \end{figure}

\begin{figure}[htbp!]
        \centering
        \includegraphics[width=0.5\textwidth]{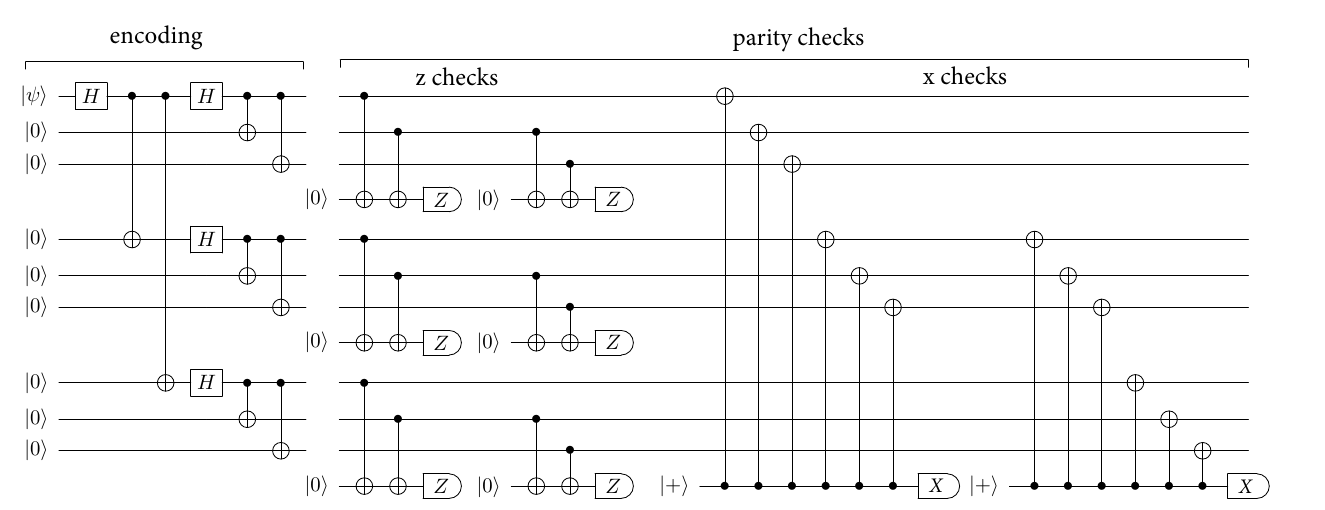}
        \caption{The encoding and the parity check circuit for the Shor code.}
        \label{fig:shor_encode}
    \end{figure}

\begin{figure}[htbp!]
        \centering
        \includegraphics[width=0.3\textwidth]{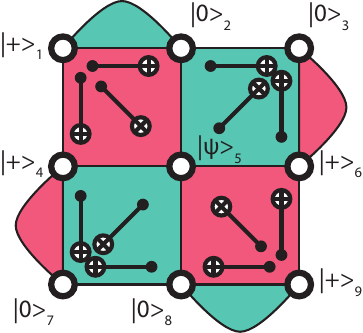}
        \caption{The encoding circuit for the Surface code. The parities are checked using similar circuits as our Steane code circuit and Shor code circuit, with one ancilla per parity check and using CNOT for the interaction between data and ancilla.}
        \label{fig:surface_encode}
    \end{figure}

    Here in Figure~\ref{fig:steane_encode}, \ref{fig:shor_encode} and \ref{fig:surface_encode}, we outline the encoding circuits and the parity check circuits we used for our codes. 
    \FloatBarrier
    \section{Multiple Rounds of Error Correction under Global Z Rotation}\label{sect:multi_same_coher}
    \subsection{Without logical twirling}
    All the codes that we have considered in this section have one logical qubit and transversal $Z$ gates. When they undergo coherent $Z$ noise, the effective logical error for a given measured syndrome $\vec{m}$ after correction will be a logical $Z$ rotation of angle $\theta_{\vec{m}}$. Hence, the effective logical error channel averaged over all the syndrome measurements is:
    \begin{align}\label{eqn:one_dir}
    \overline{\mathcal{N}}_0 = \sum_{\vec{m}} p_{\vec{m}} \supop{\overline{Z}(\theta_{\vec{m}})}
    \end{align}
    The Pauli transfer matrix of $\overline{Z}(\theta)$ is 
    \begin{align*}
    \begin{pmatrix}
    1 & 0 & 0 & 0\\  0 & \cos(\theta) & - \sin(\theta) & 0\\ 0 & \sin(\theta) & \cos(\theta) & 0\\ 0 & 0 & 0 & 1
    \end{pmatrix}
    \end{align*}
    with the eigenvalues $1, 1, e^{- i \theta}, e^{i \theta}$, and the same eigenvectors independent of $\theta$. Hence, $\overline{\mathcal{N}}_0$ will have the same eigenvectors with the eigenvalues $1, 1,  \sum_{\vec{m}} p_{\vec{m}} e^{-i\theta_{\vec{m}}}, \sum_{\vec{m}} p_{\vec{m}} e^{i\theta_{\vec{m}}}$.
    
    Hence, after $k$ round of error correction, the logical fidelity is:
    \begin{align} \label{eqn:multi_round_coher}
    F(\overline{\mathcal{N}}_0^k) &= \frac{\frac{1}{2}\Tr{\overline{\mathcal{N}}_0^k} + 1}{3} \nonumber\\ 
    &= \frac{\Re{\left(\sum_{\vec{m}} p_{\vec{m}} e^{-i\theta_{\vec{m}}}\right)^k} + 2}{3}
    \end{align}
    
    If we twirl the $Z$ noise channel, then we will have a logical dephasing channel instead. Thus $\overline{\mathcal{N}}_T$ is a diagonal matrix with eigenvalues: $1, 1-2p_{d}, 1-2p_{d}, 1$ where $p_{d}$ is the dephasing probability.
    
\begin{figure}[htbp]
        \centering
        \subfloat[]{\includegraphics[width=0.5\textwidth]{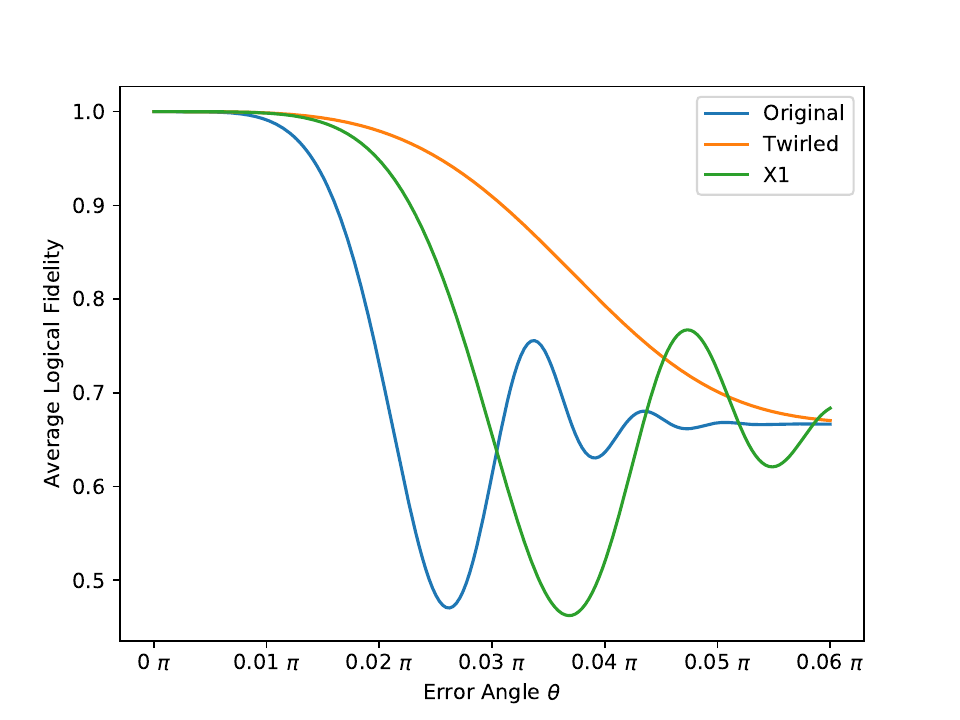}}\\
        \subfloat[]{\includegraphics[width=0.5\textwidth]{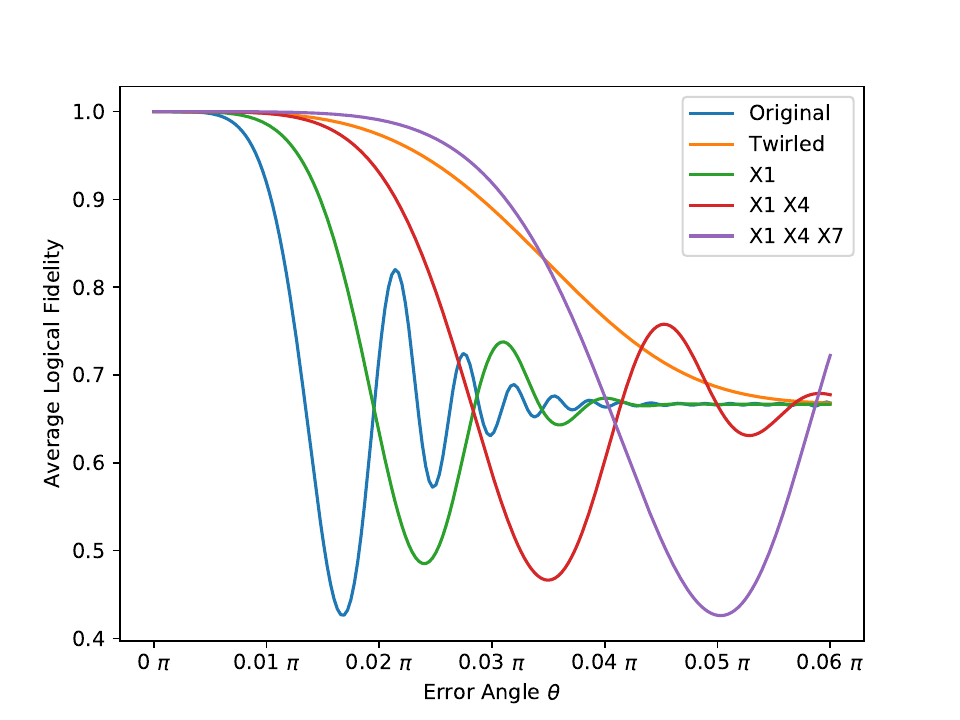}}\\
        \subfloat[]{\includegraphics[width=0.5\textwidth]{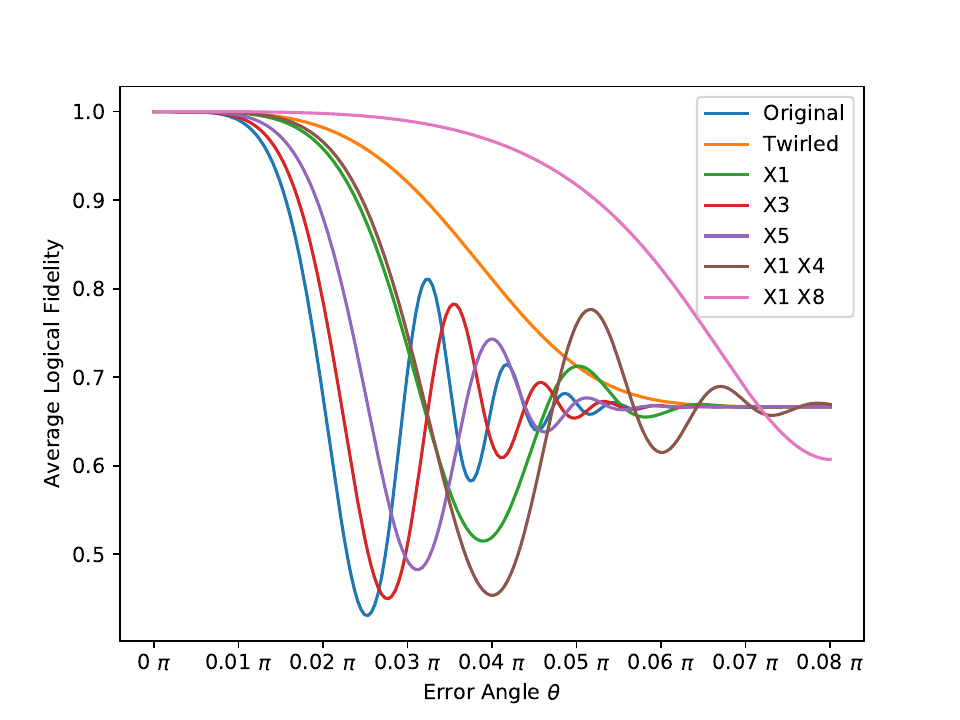}}
        \caption{Logical fidelity after 100 cycles of noise with quantum error correction using different strategies for (a) Steane code, (b)nine-qubit Shor code and (c) distance-3 surface code. Within \emph{each} round the noise is the same coherent rotation of the strength $\theta$. These figure illustrate the issue that is tackled through `logical twirling' as we explain Section~\ref{sect:multi}.}
        \label{fig:fidelity_multi}
    \end{figure}

    Hence, for $k$ round of error correction (each round undergo the same noise as before), the logical fidelity is:
    \begin{align}\label{eqn:multi_round_twirled}
    F(\overline{\mathcal{N}}_T^k) &= \frac{\frac{1}{2}\Tr{\overline{\mathcal{N}}_T^k} + 1}{3} \nonumber\\ 
    &= \frac{(1-2p_{d})^k + 2}{3}
    \end{align}
    Using these formulae, in Figure~\ref{fig:fidelity_multi} we have plotted the logical fidelity of different schemes for different codes after $100$ cycles; in each cycle, error correction follows a period of exposure to the environment which induces global $Z$ rotation with angle $\theta$. In all codes, we can see the improvements of the Pauli conjugation schemes and the twirling scheme over doing nothing in small $\theta$. In Shor code and surface code, we can see the advantage of our optimal conjugation scheme over twirling still remains for \emph{small error angle $\theta$ in each round} even after $100$ rounds. Nevertheless, we can see that in many cases twirling becomes superior to even our best Pauli conjugation after sufficient cycles have occurred; fortunately this can be entirely remedied by an adaption we term `logical twirling'.  
    \subsection{With logical twirling}
    The logically twirled version of the noise channel $\overline{\mathcal{N}}_{0}$ described in (\ref{eqn:one_dir}) is a logical dephasing channel of the form
    \begin{align}
    \overline{\mathcal{N}}_{0, LT} = \sum_{\vec{m}} p_{\vec{m}} \left(\cos[2](\frac{\phi_{\vec{m}}}{2}) \supop{\overline{I}} + \sin[2](\frac{\phi_{\vec{m}}}{2}) \supop{\overline{Z}}\right) \label{eqn:avg_two_dir}
    \end{align}
    The corresponding conjugated noise channel $\overline{\mathcal{N}}_{c}$ will be in a form similar to (\ref{eqn:one_dir}) with different $\phi_{\vec{m}}$. Applying logical twirling on top of conjugation will lead to the channel $\overline{\mathcal{N}}_{c, LT}$. Recall that the corresponding physically twirled noise channel is denoted as $\overline{\mathcal{N}}_{T}$.
    
    Our previous simulations for one round of error correction show that:
    \begin{align*}
    F(\overline{\mathcal{N}}_{c}) \geq F(\overline{\mathcal{N}}_T) \geq F(\overline{\mathcal{N}}_{0}).
    \end{align*}
    Since logical twirling will not change the logical fidelity (since the eigenvalues of the Pauli transfer matrices are not affected), we have:
    \begin{align}
    F(\overline{\mathcal{N}}_{c, LT}) \geq F(\overline{\mathcal{N}}_{T}) \geq F(\overline{\mathcal{N}}_{0, LT}). \label{eqn:one_round_compare}
    \end{align}
    in which $\overline{\mathcal{N}}_{c, LT}$, $\overline{\mathcal{N}}_{T}$ and $\overline{\mathcal{N}}_{0, LT}$ are all logical dephasing channel with different dephasing probability $p_d$. 
    
    For a single-qubit dephasing channel $\overline{\mathcal{N}}_d$ with dephasing probability $p_d$, the eigenvalues of its Pauli transfer matrix will be $\{\lambda_i\} = \{1, 1-2p_d, 1-2p_d, 1\}$. Hence, the logical fidelity of $k$ rounds of $\overline{\mathcal{N}}_{d}$ is~\cite{kimmelRobustExtractionTomographic2014, helsenSpectralQuantumTomography2019}:
    \begin{align*}\label{eqn:multi_round_fidelity}
    F(\overline{\mathcal{N}}_{d}^k) &= \frac{\Tr{\overline{\mathcal{N}}_{d}^k} + 2}{6}\\
    & = \frac{\sum_{i} \lambda_i^k + 2}{6}\\
    & = \frac{(1-2p_{d})^k + 2}{3}
    \end{align*}
    Thus we have:
    \begin{align*}
    F(\overline{\mathcal{N}}_{d}) \geq F(\overline{\mathcal{N}}_{d}') \Rightarrow F(\overline{\mathcal{N}}_{d}^k) \geq F(\overline{\mathcal{N}}_{d}'^k) \quad \forall k \in \mathbb{Z}_+
    \end{align*}
    Combining with (\ref{eqn:one_round_compare}), we then have:
    \begin{align*}
    F(\overline{\mathcal{N}}_{c, LT}^k) \geq F(\overline{\mathcal{N}}_T^k) \geq F(\overline{\mathcal{N}}_{0, LT}^k).
    \end{align*}
    for any positive integer $k$. Hence, with the help of logical twirling, the improvement of logical fidelity using Pauli conjugation over twirling (or doing nothing) with single-round of error correction in global Z rotation will indeed persist when we go to multiple rounds of error corrections.
    
    \subsection{Random walk noise model}\label{sect:random_walk}
    Up to now, we have only considered the case in which the global $Z$ rotations in each round of error correction are rotations of the same angle in the same direction. In practice, for such a noise model, all we need to do is flip all the qubits right in the middle of the whole process which flips the direction of the rotation and cancels the coherent error. This is just a simple case of dynamical decoupling.
    
    It may be interesting to look at the other extreme in which the error channel is a random walk. Within each round of error correction, there is a equal probability of positive or negative rotation of angle $\theta$: $N(\pm \theta) = e^{\pm i\theta \sum_j Z_j}$. 
    
    A a global $Z$ rotation $e^{- i\theta \sum_j Z_j}$ will lead to an effective logical error channel as described in (\ref{eqn:one_dir}):
    \begin{align*}
    \overline{\mathcal{N}}_{0} = \sum_{\vec{m}} p_{\vec{m}} \supop{\overline{Z}(\phi_{\vec{m}})}
    \end{align*}
    When the sign of rotation of the physical error $\theta$ is flipped, the sign of the logical rotation $\phi_{\vec{m}}$ will also be flipped for all the codes that we are considering (see Appendix~\ref{sect:sign_log_rot}). For each time step, since we have equal probabilities of positive and negative physical rotations, we also have equal probabilities of positive and negative logical rotations, leading to the effective logical channel:
    \begin{align*}
    \overline{\mathcal{N}}_{0, \pm} = \sum_{\vec{m}} p_{\vec{m}} \left(\cos[2](\frac{\phi_{\vec{m}}}{2}) \supop{\overline{I}} + \sin[2](\frac{\phi_{\vec{m}}}{2}) \supop{\overline{Z}}\right)
    \end{align*}
    which is just the logically twirled channel $\overline{\mathcal{N}}_{0, LT}$. Hence, for such a random walk noise model, the logical channel is already logically twirled and we just need to apply conjugation to it to reduce the effect of the noise.

 \section{Conjugating High Frequency Noise}\label{sect:high_freq}
    For a coherent noise:
    \begin{align*}
    U(\theta) &= e^{-i H t},
    \end{align*}
    the Hamiltonian $H$ can be broken down into its Pauli basis $\mathbb{G}_H$:
    \begin{align*}
    H = \sum_{g_i \in \mathbb{G}_H} \beta_ig_i
    \end{align*}
    Note that $\beta_i$ are real since $H$ is Hermitian.
    
    Now we define the magnitude of $H$ to be $E$, and the normalised version of $H$ to be $h$ where:
    \begin{align}
    E &= \sqrt{\sum_i \beta_i^2}\\
    h &= \frac{H}{E} = \sum_{{g}_i \in G_{{H}}} \frac{\beta_i}{E} {g}_i = \sum_{{g}_i \in G_{{H}}} \alpha_i {g}_i \label{eqn:h_decomp}
    \end{align}
    for $\alpha_i = \frac{\beta_i}{E}$ and $\sum_i \alpha_i^2 = 1$.
    
    Now the evolution operator is just:
    \begin{align*}
    {U}(t)& = e^{-i{H} t}  = e^{-i{h} Et}\\
    {U}(\theta) &= e^{-i\theta{h}}
    \end{align*}
    with $\theta = Et$. 
    
    Suppose our noise channel is some high frequency noise that is only coherent for a very short amount of time $\delta t$, resulting in a rotation angle of $\epsilon = E \delta t$:
    \begin{align*}
    {U}(\pm \epsilon) &= e^{\pm i \epsilon {h} }\\
    & \approx {I} \pm i\epsilon{h} - \frac{\epsilon^2}{2} {h}^2\\
    & = {I} - \frac{\epsilon^2}{2} {h}^2 \pm i\epsilon{h} 
    \end{align*}
    Within each time period $\delta t$, the coherent noise will have a $50$-$50$ chance for rotations in the positive and negative directions, just like a random walk. Thus the effective channel over a time period $\delta t$ is just:
    \begin{align}
    \mathcal{U}_{\epsilon}(\rho) &= \frac{1}{2}{U}(\epsilon) \rho {U}^\dagger(\epsilon) + \frac{1}{2} {U}(- \epsilon) \rho {U}^\dagger(- \epsilon)\nonumber\\
    &=\left({I} - \frac{\epsilon^2}{2} {h}^2 \right) \rho \left({I} - \frac{\epsilon^2}{2} {h}^2 \right) + \epsilon^2{h} \rho {h}\label{eqn:fluc_channel}
    \end{align}
    which in the Pauli transfer matrix formalism is just:
    \begin{align*}
    \mathcal{U}_{\epsilon}\pket{\rho}= \left(\supop{{I} - \frac{\epsilon^2}{2} {h}^2 }  + \epsilon^2 \supop{h} \right)\pket{\rho}
    \end{align*}
    Composing $N$ of such channels together we have:
    \begin{align*}
    \mathcal{U}_{\epsilon}^N\pket{\rho} & = \left(\supop{{I} - \frac{\epsilon^2}{2} {h}^2 }  + \epsilon^2 \supop{h} \right)^N\pket{\rho}\\
    & \approx \left(\left(\supop{{I} - \frac{\epsilon^2}{2} {h}^2 }\right)^N  + N \epsilon^2 \supop{h}\left(\supop{{I} - \frac{\epsilon^2}{2} {h}^2 }\right)^{N-1}\right)\pket{\rho}\\
    & = \left(\supop{\left({I} - \frac{\epsilon^2}{2} {h}^2 \right)^N}  + N \epsilon^2 \supop{h}\supop{\left({I} - \frac{\epsilon^2}{2} {h}^2 \right)^{N-1}}\right)\pket{\rho}\\
    & \approx \left(\supop{{I} - \frac{N\epsilon^2}{2} {h}^2}  + N \epsilon^2 \supop{h}\right)\pket{\rho}\\
    & = \mathcal{U}_{\sqrt{N}\epsilon}\pket{\rho}
    \end{align*}
    Now if $H$ (and thus $h$) contains coherent superposition of multiple Pauli components, then as discussed in Section~\ref{sect:mech_conj}, conjugation can be used to improve the logical fidelity of the channel by changing the way these components interfere. In particular, if a conjugation scheme works for the channel $\mathcal{U}_{\epsilon}$, then the same scheme should also work for the composite channel $\mathcal{U}_{\epsilon}^N \approx \mathcal{U}_{\sqrt{N}\epsilon}$ since their Pauli components interfere in similar ways (as can be seen from their similar structural dependence on $h$). In the case of the global $Z$ rotation that we considered in Section~\ref{sect:Z_noise_sim}, we have discussed why conjugation would work for a single step of the random walk channel $\mathcal{U}_{\epsilon}$ in Appendix~\ref{sect:random_walk} (in which the channel is denoted as $\overline{\mathcal{N}}_{0, \pm}$). Hence, by the arguments above, the same conjugation scheme will also work for the composite channel which corresponds to high frequency global $Z$ noise.

    \section{Multi-round Twirling Set Reduction}\label{sect:multi_reduction}
    The effective error channel with $K$ rounds of twirling is:
    \begin{align*}
    (\overline{\mathcal{N}}_{TK})_{G, G'} &= \pbra{\Pi_{\vec{0}}\overline{G}} \mathcal{R} \left[\prod_{k=1}^{K}\mathcal{T}(\mathcal{N})\right] \pket{\Pi_{\vec{0}}\overline{G}'}
    \end{align*}
    The argument about structure of noise (Section~\ref{sect:noise_sym}) can still be applied to the twirling within each individual rounds here, giving us a smaller set of twirling generators $\widetilde{\mathbb{W}}$, from which we can obtained a reduced twirling set $\mathbb{W}$:
    \begin{align*}
    (\overline{\mathcal{N}}_{TK})_{G, G'} &= \frac{1}{\abs{\mathbb{W}}^K}\sum_{\vec{W} \in \mathbb{W}^K}\pbra{\Pi_{\vec{0}}\overline{G}} \mathcal{R} \prod_{k=1}^{K}{\supop{W}}_k \mathcal{N} {\supop{W}}_k \pket{\Pi_{\vec{0}}\overline{G}'}
    \end{align*}
    Since we are summing all possible $\vec{W}$ and the twirling set is a group on super-operator composition, we can do the following change of variables: ${\supop{W}}_k{\supop{W}}_{k + 1} \Rightarrow {\supop{W}}_{k + 1}$, which gives:
    \begin{align*}
    &\quad (\overline{\mathcal{N}}_{TK})_{G, G'} \nonumber\\
    &=  \frac{1}{\abs{\mathbb{W}}^K}\sum_{\vec{W} \in \mathbb{W}^K} \pbra{\Pi_{\vec{0}}\overline{G}} \mathcal{R} \left(\prod_{k=1}^{K}{\supop{W}}_k\right) \left(\prod_{k=K}^{1} \mathcal{N} {\supop{W}}_k\right)\pket{\Pi_{\vec{0}}\overline{G}'}
    \end{align*}
    In this form, our arguments about interaction of twirling with the code space in Section~\ref{sect:remove_stb} can be applied to the outermost twirling set, obtaining a reduced twirling set $\mathbb{W}_1$. Hence, we have 
    \begin{align*}
    &\quad (\overline{\mathcal{N}}_{TK})_{G, G'} =  \frac{1}{\abs{\mathbb{W}_1}\abs{\mathbb{W}}^{K-1}}\\
    & \quad \times \sum_{\vec{W} \in \mathbb{W}_1 \times \mathbb{W}^{K - 1}} \pbra{\Pi_{\vec{0}}\overline{G}} \mathcal{R} \left(\prod_{k=1}^{K}{\supop{W}}_k\right) \left(\prod_{k=K}^{1} \mathcal{N} {\supop{W}}_k\right)\pket{\Pi_{\vec{0}}\overline{G}'}
    \end{align*}
    Similar arguments to Section~\ref{sect:perm_sym} can be made about the symmetries in both noise and code. However, rather than proving the equivalence of using two different Pauli operators in conjugation, we now will prove the equivalence of using two different \emph{sets} of Pauli operators in conjugation: i.e. after find the symmetry $U$, we can say a Pauli conjugation set $\vec{W}'$ is equivalent to $\vec{W}$ when $U\vec{W}U^\dagger = \vec{W}'$. Here $U\vec{W}U^\dagger$ is defined as:
    \begin{align*}
    U\vec{W}U^\dagger = (UW_1U^\dagger, UW_2U^\dagger, \cdots, UW_KU^\dagger)
    \end{align*}
    Note that this is not a simple tensor product of the single round case. For example, if $W_1$ equivalent to $W_1'$ due to symmetry $U$ and $W_2$ equivalent to $W_2'$ due to another symmetry $U'$, this does not means that $\vec{W} = (W_1, W_2)$ is equivalent to $\vec{W}' = (W_1', W_2')$ since the two elements are related by different symmetry: $U\vec{W}U^\dagger\neq \vec{W} \neq U'\vec{W}U'^\dagger$.
    
    \section{Notation and Definition}
    \begin{itemize}
        \item[$\supop{\quad}$:] Super-operators. e.g. $\supop{A}(\rho) = A\rho A^\dagger$.
        \item[$\supsupop{\quad}$:] Super-super-operators. e.g. $\supsupop{A} (\mathcal{N})= \supop{A}\mathcal{N} \supop{A^\dagger}$.
        \item[$\eta$:] Commutator. $AB = \eta(A, B)BA$ .
        \item[$\widetilde{\quad}$:] Generating set. Note that $\widetilde{A}$ means that $A$ can be generated from $\widetilde{A}$, but does not means that $A$ is the \textbf{complete }set of elements that can be generated from $\widetilde{A}$. In our paper, all the composition are carried out ignoring the irrelevant phase factor of the Pauli operators. 
        \item[$\mathbb{G}$:] The Pauli set. It is \emph{not} the Pauli group since we are ignoring all the phase factors.
        \item [$\mathbb{W}$:] The twirling set.
    \end{itemize}
%    \newpage
    %\bibliographystyle{natbib}
%    \bibliography{ref}

\begin{thebibliography}{39}%
\makeatletter
\providecommand \@ifxundefined [1]{%
 \@ifx{#1\undefined}
}%
\providecommand \@ifnum [1]{%
 \ifnum #1\expandafter \@firstoftwo
 \else \expandafter \@secondoftwo
 \fi
}%
\providecommand \@ifx [1]{%
 \ifx #1\expandafter \@firstoftwo
 \else \expandafter \@secondoftwo
 \fi
}%
\providecommand \natexlab [1]{#1}%
\providecommand \enquote  [1]{``#1''}%
\providecommand \bibnamefont  [1]{#1}%
\providecommand \bibfnamefont [1]{#1}%
\providecommand \citenamefont [1]{#1}%
\providecommand \href@noop [0]{\@secondoftwo}%
\providecommand \href [0]{\begingroup \@sanitize@url \@href}%
\providecommand \@href[1]{\@@startlink{#1}\@@href}%
\providecommand \@@href[1]{\endgroup#1\@@endlink}%
\providecommand \@sanitize@url [0]{\catcode `\\12\catcode `\$12\catcode
  `\&12\catcode `\#12\catcode `\^12\catcode `\_12\catcode `\%12\relax}%
\providecommand \@@startlink[1]{}%
\providecommand \@@endlink[0]{}%
\providecommand \url  [0]{\begingroup\@sanitize@url \@url }%
\providecommand \@url [1]{\endgroup\@href {#1}{\urlprefix }}%
\providecommand \urlprefix  [0]{URL }%
\providecommand \Eprint [0]{\href }%
\providecommand \doibase [0]{http://dx.doi.org/}%
\providecommand \selectlanguage [0]{\@gobble}%
\providecommand \bibinfo  [0]{\@secondoftwo}%
\providecommand \bibfield  [0]{\@secondoftwo}%
\providecommand \translation [1]{[#1]}%
\providecommand \BibitemOpen [0]{}%
\providecommand \bibitemStop [0]{}%
\providecommand \bibitemNoStop [0]{.\EOS\space}%
\providecommand \EOS [0]{\spacefactor3000\relax}%
\providecommand \BibitemShut  [1]{\csname bibitem#1\endcsname}%
\let\auto@bib@innerbib\@empty
%</preamble>
\bibitem [{\citenamefont {Aharonov}\ and\ \citenamefont
  {{Ben-Or}}(1997)}]{aharonovFaulttolerantQuantumComputation1997}%
  \BibitemOpen
  \bibfield  {author} {\bibinfo {author} {\bibfnamefont {D.}~\bibnamefont
  {Aharonov}}\ and\ \bibinfo {author} {\bibfnamefont {M.}~\bibnamefont
  {{Ben-Or}}},\ }\bibfield  {title} {\enquote {\bibinfo {title} {Fault-tolerant
  {{Quantum Computation}} with {{Constant Error}}},}\ }in\ \href {\doibase
  10.1145/258533.258579} {\emph {\bibinfo {booktitle} {Proceedings of the
  {{Twenty}}-Ninth {{Annual ACM Symposium}} on {{Theory}} of {{Computing}}}}},\
  \bibinfo {series and number} {{{STOC}} '97}\ (\bibinfo  {publisher} {{ACM}},\
  \bibinfo {address} {{New York, NY, USA}},\ \bibinfo {year} {1997})\ pp.\
  \bibinfo {pages} {176--188}\BibitemShut {NoStop}%
\bibitem [{\citenamefont {Knill}\ \emph {et~al.}(1998)\citenamefont {Knill},
  \citenamefont {Laflamme},\ and\ \citenamefont
  {Zurek}}]{knillResilientQuantumComputation1998}%
  \BibitemOpen
  \bibfield  {author} {\bibinfo {author} {\bibfnamefont {Emanuel}\ \bibnamefont
  {Knill}}, \bibinfo {author} {\bibfnamefont {Raymond}\ \bibnamefont
  {Laflamme}}, \ and\ \bibinfo {author} {\bibfnamefont {Wojciech~H.}\
  \bibnamefont {Zurek}},\ }\bibfield  {title} {\enquote {\bibinfo {title}
  {Resilient {{Quantum Computation}}},}\ }\href {\doibase
  10.1126/science.279.5349.342} {\bibfield  {journal} {\bibinfo  {journal}
  {Science}\ }\textbf {\bibinfo {volume} {279}},\ \bibinfo {pages} {342--345}
  (\bibinfo {year} {1998})}\BibitemShut {NoStop}%
\bibitem [{\citenamefont {Aliferis}\ \emph {et~al.}(2006)\citenamefont
  {Aliferis}, \citenamefont {Gottesman},\ and\ \citenamefont
  {Preskill}}]{aliferisQuantumAccuracyThreshold2006}%
  \BibitemOpen
  \bibfield  {author} {\bibinfo {author} {\bibfnamefont {Panos}\ \bibnamefont
  {Aliferis}}, \bibinfo {author} {\bibfnamefont {Daniel}\ \bibnamefont
  {Gottesman}}, \ and\ \bibinfo {author} {\bibfnamefont {John}\ \bibnamefont
  {Preskill}},\ }\bibfield  {title} {\enquote {\bibinfo {title} {Quantum
  {{Accuracy Threshold}} for {{Concatenated Distance}}-3 {{Codes}}},}\ }\href
  {http://dl.acm.org/citation.cfm?id=2011665.2011666} {\bibfield  {journal}
  {\bibinfo  {journal} {Quantum Info. Comput.}\ }\textbf {\bibinfo {volume}
  {6}},\ \bibinfo {pages} {97--165} (\bibinfo {year} {2006})}\BibitemShut
  {NoStop}%
\bibitem [{\citenamefont {Sanders}\ \emph {et~al.}(2015)\citenamefont
  {Sanders}, \citenamefont {Wallman},\ and\ \citenamefont
  {Sanders}}]{sandersBoundingQuantumGate2015}%
  \BibitemOpen
  \bibfield  {author} {\bibinfo {author} {\bibfnamefont {Yuval~R.}\
  \bibnamefont {Sanders}}, \bibinfo {author} {\bibfnamefont {Joel~J.}\
  \bibnamefont {Wallman}}, \ and\ \bibinfo {author} {\bibfnamefont {Barry~C.}\
  \bibnamefont {Sanders}},\ }\bibfield  {title} {\enquote {\bibinfo {title}
  {Bounding quantum gate error rate based on reported average fidelity},}\
  }\href {\doibase 10.1088/1367-2630/18/1/012002} {\bibfield  {journal}
  {\bibinfo  {journal} {New Journal of Physics}\ }\textbf {\bibinfo {volume}
  {18}},\ \bibinfo {pages} {012002} (\bibinfo {year} {2015})}\BibitemShut
  {NoStop}%
\bibitem [{\citenamefont {Guti{\'e}rrez}\ and\ \citenamefont
  {Brown}(2015)}]{gutierrezComparisonQuantumErrorcorrection2015}%
  \BibitemOpen
  \bibfield  {author} {\bibinfo {author} {\bibfnamefont {Mauricio}\
  \bibnamefont {Guti{\'e}rrez}}\ and\ \bibinfo {author} {\bibfnamefont
  {Kenneth~R.}\ \bibnamefont {Brown}},\ }\bibfield  {title} {\enquote {\bibinfo
  {title} {Comparison of a quantum error-correction threshold for exact and
  approximate errors},}\ }\href {\doibase 10.1103/PhysRevA.91.022335}
  {\bibfield  {journal} {\bibinfo  {journal} {Physical Review A}\ }\textbf
  {\bibinfo {volume} {91}},\ \bibinfo {pages} {022335} (\bibinfo {year}
  {2015})}\BibitemShut {NoStop}%
\bibitem [{\citenamefont {Kueng}\ \emph {et~al.}(2016)\citenamefont {Kueng},
  \citenamefont {Long}, \citenamefont {Doherty},\ and\ \citenamefont
  {Flammia}}]{kuengComparingExperimentsFaultTolerance2016a}%
  \BibitemOpen
  \bibfield  {author} {\bibinfo {author} {\bibfnamefont {Richard}\ \bibnamefont
  {Kueng}}, \bibinfo {author} {\bibfnamefont {David~M.}\ \bibnamefont {Long}},
  \bibinfo {author} {\bibfnamefont {Andrew~C.}\ \bibnamefont {Doherty}}, \ and\
  \bibinfo {author} {\bibfnamefont {Steven~T.}\ \bibnamefont {Flammia}},\
  }\bibfield  {title} {\enquote {\bibinfo {title} {Comparing {{Experiments}} to
  the {{Fault}}-{{Tolerance Threshold}}},}\ }\href {\doibase
  10.1103/PhysRevLett.117.170502} {\bibfield  {journal} {\bibinfo  {journal}
  {Physical Review Letters}\ }\textbf {\bibinfo {volume} {117}},\ \bibinfo
  {pages} {170502} (\bibinfo {year} {2016})}\BibitemShut {NoStop}%
\bibitem [{\citenamefont {Bravyi}\ \emph {et~al.}(2018)\citenamefont {Bravyi},
  \citenamefont {Englbrecht}, \citenamefont {K{\"o}nig},\ and\ \citenamefont
  {Peard}}]{bravyiCorrectingCoherentErrors2018}%
  \BibitemOpen
  \bibfield  {author} {\bibinfo {author} {\bibfnamefont {Sergey}\ \bibnamefont
  {Bravyi}}, \bibinfo {author} {\bibfnamefont {Matthias}\ \bibnamefont
  {Englbrecht}}, \bibinfo {author} {\bibfnamefont {Robert}\ \bibnamefont
  {K{\"o}nig}}, \ and\ \bibinfo {author} {\bibfnamefont {Nolan}\ \bibnamefont
  {Peard}},\ }\bibfield  {title} {\enquote {\bibinfo {title} {Correcting
  coherent errors with surface codes},}\ }\href {\doibase
  10.1038/s41534-018-0106-y} {\bibfield  {journal} {\bibinfo  {journal} {npj
  Quantum Information}\ }\textbf {\bibinfo {volume} {4}},\ \bibinfo {pages}
  {55} (\bibinfo {year} {2018})}\BibitemShut {NoStop}%
\bibitem [{\citenamefont {Greenbaum}\ and\ \citenamefont
  {Dutton}(2018)}]{greenbaumModelingCoherentErrors2018}%
  \BibitemOpen
  \bibfield  {author} {\bibinfo {author} {\bibfnamefont {Daniel}\ \bibnamefont
  {Greenbaum}}\ and\ \bibinfo {author} {\bibfnamefont {Zachary}\ \bibnamefont
  {Dutton}},\ }\bibfield  {title} {\enquote {\bibinfo {title} {Modeling
  coherent errors in quantum error correction},}\ }\href {\doibase
  10.1088/2058-9565/aa9a06} {\bibfield  {journal} {\bibinfo  {journal} {Quantum
  Science and Technology}\ }\textbf {\bibinfo {volume} {3}},\ \bibinfo {pages}
  {015007} (\bibinfo {year} {2018})}\BibitemShut {NoStop}%
\bibitem [{\citenamefont {Iyer}\ and\ \citenamefont
  {Poulin}(2018)}]{iyerSmallQuantumComputer2018}%
  \BibitemOpen
  \bibfield  {author} {\bibinfo {author} {\bibfnamefont {Pavithran}\
  \bibnamefont {Iyer}}\ and\ \bibinfo {author} {\bibfnamefont {David}\
  \bibnamefont {Poulin}},\ }\bibfield  {title} {\enquote {\bibinfo {title} {A
  small quantum computer is needed to optimize fault-tolerant protocols},}\
  }\href {\doibase 10.1088/2058-9565/aab73c} {\bibfield  {journal} {\bibinfo
  {journal} {Quantum Science and Technology}\ }\textbf {\bibinfo {volume}
  {3}},\ \bibinfo {pages} {030504} (\bibinfo {year} {2018})}\BibitemShut
  {NoStop}%
\bibitem [{\citenamefont {Huang}\ \emph {et~al.}(2019)\citenamefont {Huang},
  \citenamefont {Doherty},\ and\ \citenamefont
  {Flammia}}]{huangPerformanceQuantumError2019}%
  \BibitemOpen
  \bibfield  {author} {\bibinfo {author} {\bibfnamefont {Eric}\ \bibnamefont
  {Huang}}, \bibinfo {author} {\bibfnamefont {Andrew~C.}\ \bibnamefont
  {Doherty}}, \ and\ \bibinfo {author} {\bibfnamefont {Steven}\ \bibnamefont
  {Flammia}},\ }\bibfield  {title} {\enquote {\bibinfo {title} {Performance of
  quantum error correction with coherent errors},}\ }\href {\doibase
  10.1103/PhysRevA.99.022313} {\bibfield  {journal} {\bibinfo  {journal}
  {Physical Review A}\ }\textbf {\bibinfo {volume} {99}},\ \bibinfo {pages}
  {022313} (\bibinfo {year} {2019})}\BibitemShut {NoStop}%
\bibitem [{\citenamefont {Lidar}(2014)}]{lidarReviewDecoherenceFree2014}%
  \BibitemOpen
  \bibfield  {author} {\bibinfo {author} {\bibfnamefont {Daniel~A.}\
  \bibnamefont {Lidar}},\ }\bibfield  {title} {\enquote {\bibinfo {title}
  {Review of {{Decoherence Free Subspaces}}, {{Noiseless Subsystems}}, and
  {{Dynamical Decoupling}}},}\ }\href {\doibase 10.1002/9781118742631}
  {\bibfield  {journal} {\bibinfo  {journal} {arXiv:1208.5791 [cond-mat,
  physics:physics, physics:quant-ph]}\ } (\bibinfo {year} {2014}),\
  10.1002/9781118742631}\BibitemShut {NoStop}%
\bibitem [{\citenamefont {Suter}\ and\ \citenamefont
  {{\'A}lvarez}(2016)}]{suterColloquiumProtectingQuantum2016}%
  \BibitemOpen
  \bibfield  {author} {\bibinfo {author} {\bibfnamefont {Dieter}\ \bibnamefont
  {Suter}}\ and\ \bibinfo {author} {\bibfnamefont {Gonzalo~A.}\ \bibnamefont
  {{\'A}lvarez}},\ }\bibfield  {title} {\enquote {\bibinfo {title} {Colloquium:
  {{Protecting}} quantum information against environmental noise},}\ }\href
  {\doibase 10.1103/RevModPhys.88.041001} {\bibfield  {journal} {\bibinfo
  {journal} {Reviews of Modern Physics}\ }\textbf {\bibinfo {volume} {88}},\
  \bibinfo {pages} {041001} (\bibinfo {year} {2016})}\BibitemShut {NoStop}%
\bibitem [{\citenamefont {Beale}\ \emph {et~al.}(2018)\citenamefont {Beale},
  \citenamefont {Wallman}, \citenamefont {Guti{\'e}rrez}, \citenamefont
  {Brown},\ and\ \citenamefont {Laflamme}}]{bealeQuantumErrorCorrection2018}%
  \BibitemOpen
  \bibfield  {author} {\bibinfo {author} {\bibfnamefont {Stefanie~J.}\
  \bibnamefont {Beale}}, \bibinfo {author} {\bibfnamefont {Joel~J.}\
  \bibnamefont {Wallman}}, \bibinfo {author} {\bibfnamefont {Mauricio}\
  \bibnamefont {Guti{\'e}rrez}}, \bibinfo {author} {\bibfnamefont {Kenneth~R.}\
  \bibnamefont {Brown}}, \ and\ \bibinfo {author} {\bibfnamefont {Raymond}\
  \bibnamefont {Laflamme}},\ }\bibfield  {title} {\enquote {\bibinfo {title}
  {Quantum {{Error Correction Decoheres Noise}}},}\ }\href {\doibase
  10.1103/PhysRevLett.121.190501} {\bibfield  {journal} {\bibinfo  {journal}
  {Physical Review Letters}\ }\textbf {\bibinfo {volume} {121}},\ \bibinfo
  {pages} {190501} (\bibinfo {year} {2018})}\BibitemShut {NoStop}%
\bibitem [{\citenamefont {Chamberland}\ \emph {et~al.}(2017)\citenamefont
  {Chamberland}, \citenamefont {Wallman}, \citenamefont {Beale},\ and\
  \citenamefont {Laflamme}}]{chamberlandHardDecodingAlgorithm2017}%
  \BibitemOpen
  \bibfield  {author} {\bibinfo {author} {\bibfnamefont {Christopher}\
  \bibnamefont {Chamberland}}, \bibinfo {author} {\bibfnamefont {Joel}\
  \bibnamefont {Wallman}}, \bibinfo {author} {\bibfnamefont {Stefanie}\
  \bibnamefont {Beale}}, \ and\ \bibinfo {author} {\bibfnamefont {Raymond}\
  \bibnamefont {Laflamme}},\ }\bibfield  {title} {\enquote {\bibinfo {title}
  {Hard decoding algorithm for optimizing thresholds under general
  {{Markovian}} noise},}\ }\href {\doibase 10.1103/PhysRevA.95.042332}
  {\bibfield  {journal} {\bibinfo  {journal} {Physical Review A}\ }\textbf
  {\bibinfo {volume} {95}},\ \bibinfo {pages} {042332} (\bibinfo {year}
  {2017})}\BibitemShut {NoStop}%
\bibitem [{\citenamefont {Debroy}\ \emph {et~al.}(2018)\citenamefont {Debroy},
  \citenamefont {Li}, \citenamefont {Newman},\ and\ \citenamefont
  {Brown}}]{debroyStabilizerSlicingCoherent2018}%
  \BibitemOpen
  \bibfield  {author} {\bibinfo {author} {\bibfnamefont {Dripto~M.}\
  \bibnamefont {Debroy}}, \bibinfo {author} {\bibfnamefont {Muyuan}\
  \bibnamefont {Li}}, \bibinfo {author} {\bibfnamefont {Michael}\ \bibnamefont
  {Newman}}, \ and\ \bibinfo {author} {\bibfnamefont {Kenneth~R.}\ \bibnamefont
  {Brown}},\ }\bibfield  {title} {\enquote {\bibinfo {title} {Stabilizer
  {{Slicing}}: {{Coherent Error Cancellations}} in {{Low}}-{{Density
  Parity}}-{{Check Stabilizer Codes}}},}\ }\href {\doibase
  10.1103/PhysRevLett.121.250502} {\bibfield  {journal} {\bibinfo  {journal}
  {Physical Review Letters}\ }\textbf {\bibinfo {volume} {121}},\ \bibinfo
  {pages} {250502} (\bibinfo {year} {2018})}\BibitemShut {NoStop}%
\bibitem [{\citenamefont {Bennett}\ \emph
  {et~al.}(1996{\natexlab{a}})\citenamefont {Bennett}, \citenamefont
  {Brassard}, \citenamefont {Popescu}, \citenamefont {Schumacher},
  \citenamefont {Smolin},\ and\ \citenamefont
  {Wootters}}]{bennettPurificationNoisyEntanglement1996}%
  \BibitemOpen
  \bibfield  {author} {\bibinfo {author} {\bibfnamefont {Charles~H.}\
  \bibnamefont {Bennett}}, \bibinfo {author} {\bibfnamefont {Gilles}\
  \bibnamefont {Brassard}}, \bibinfo {author} {\bibfnamefont {Sandu}\
  \bibnamefont {Popescu}}, \bibinfo {author} {\bibfnamefont {Benjamin}\
  \bibnamefont {Schumacher}}, \bibinfo {author} {\bibfnamefont {John~A.}\
  \bibnamefont {Smolin}}, \ and\ \bibinfo {author} {\bibfnamefont {William~K.}\
  \bibnamefont {Wootters}},\ }\bibfield  {title} {\enquote {\bibinfo {title}
  {Purification of {{Noisy Entanglement}} and {{Faithful Teleportation}} via
  {{Noisy Channels}}},}\ }\href {\doibase 10.1103/PhysRevLett.76.722}
  {\bibfield  {journal} {\bibinfo  {journal} {Physical Review Letters}\
  }\textbf {\bibinfo {volume} {76}},\ \bibinfo {pages} {722--725} (\bibinfo
  {year} {1996}{\natexlab{a}})}\BibitemShut {NoStop}%
\bibitem [{\citenamefont {Bennett}\ \emph
  {et~al.}(1996{\natexlab{b}})\citenamefont {Bennett}, \citenamefont
  {DiVincenzo}, \citenamefont {Smolin},\ and\ \citenamefont
  {Wootters}}]{bennettMixedstateEntanglementQuantum1996}%
  \BibitemOpen
  \bibfield  {author} {\bibinfo {author} {\bibfnamefont {Charles~H.}\
  \bibnamefont {Bennett}}, \bibinfo {author} {\bibfnamefont {David~P.}\
  \bibnamefont {DiVincenzo}}, \bibinfo {author} {\bibfnamefont {John~A.}\
  \bibnamefont {Smolin}}, \ and\ \bibinfo {author} {\bibfnamefont {William~K.}\
  \bibnamefont {Wootters}},\ }\bibfield  {title} {\enquote {\bibinfo {title}
  {Mixed-state entanglement and quantum error correction},}\ }\href {\doibase
  10.1103/PhysRevA.54.3824} {\bibfield  {journal} {\bibinfo  {journal}
  {Physical Review A}\ }\textbf {\bibinfo {volume} {54}},\ \bibinfo {pages}
  {3824--3851} (\bibinfo {year} {1996}{\natexlab{b}})}\BibitemShut {NoStop}%
\bibitem [{\citenamefont {Knill}\ \emph {et~al.}(2008)\citenamefont {Knill},
  \citenamefont {Leibfried}, \citenamefont {Reichle}, \citenamefont {Britton},
  \citenamefont {Blakestad}, \citenamefont {Jost}, \citenamefont {Langer},
  \citenamefont {Ozeri}, \citenamefont {Seidelin},\ and\ \citenamefont
  {Wineland}}]{knillRandomizedBenchmarkingQuantum2008}%
  \BibitemOpen
  \bibfield  {author} {\bibinfo {author} {\bibfnamefont {E.}~\bibnamefont
  {Knill}}, \bibinfo {author} {\bibfnamefont {D.}~\bibnamefont {Leibfried}},
  \bibinfo {author} {\bibfnamefont {R.}~\bibnamefont {Reichle}}, \bibinfo
  {author} {\bibfnamefont {J.}~\bibnamefont {Britton}}, \bibinfo {author}
  {\bibfnamefont {R.~B.}\ \bibnamefont {Blakestad}}, \bibinfo {author}
  {\bibfnamefont {J.~D.}\ \bibnamefont {Jost}}, \bibinfo {author}
  {\bibfnamefont {C.}~\bibnamefont {Langer}}, \bibinfo {author} {\bibfnamefont
  {R.}~\bibnamefont {Ozeri}}, \bibinfo {author} {\bibfnamefont
  {S.}~\bibnamefont {Seidelin}}, \ and\ \bibinfo {author} {\bibfnamefont
  {D.~J.}\ \bibnamefont {Wineland}},\ }\bibfield  {title} {\enquote {\bibinfo
  {title} {Randomized benchmarking of quantum gates},}\ }\href {\doibase
  10.1103/PhysRevA.77.012307} {\bibfield  {journal} {\bibinfo  {journal}
  {Physical Review A}\ }\textbf {\bibinfo {volume} {77}},\ \bibinfo {pages}
  {012307} (\bibinfo {year} {2008})}\BibitemShut {NoStop}%
\bibitem [{\citenamefont {Emerson}\ \emph {et~al.}(2007)\citenamefont
  {Emerson}, \citenamefont {Silva}, \citenamefont {Moussa}, \citenamefont
  {Ryan}, \citenamefont {Laforest}, \citenamefont {Baugh}, \citenamefont
  {Cory},\ and\ \citenamefont
  {Laflamme}}]{emersonSymmetrizedCharacterizationNoisy2007}%
  \BibitemOpen
  \bibfield  {author} {\bibinfo {author} {\bibfnamefont {J.}~\bibnamefont
  {Emerson}}, \bibinfo {author} {\bibfnamefont {M.}~\bibnamefont {Silva}},
  \bibinfo {author} {\bibfnamefont {O.}~\bibnamefont {Moussa}}, \bibinfo
  {author} {\bibfnamefont {C.}~\bibnamefont {Ryan}}, \bibinfo {author}
  {\bibfnamefont {M.}~\bibnamefont {Laforest}}, \bibinfo {author}
  {\bibfnamefont {J.}~\bibnamefont {Baugh}}, \bibinfo {author} {\bibfnamefont
  {D.~G.}\ \bibnamefont {Cory}}, \ and\ \bibinfo {author} {\bibfnamefont
  {R.}~\bibnamefont {Laflamme}},\ }\bibfield  {title} {\enquote {\bibinfo
  {title} {Symmetrized {{Characterization}} of {{Noisy Quantum Processes}}},}\
  }\href {\doibase 10.1126/science.1145699} {\bibfield  {journal} {\bibinfo
  {journal} {Science}\ }\textbf {\bibinfo {volume} {317}},\ \bibinfo {pages}
  {1893--1896} (\bibinfo {year} {2007})}\BibitemShut {NoStop}%
\bibitem [{\citenamefont
  {Greenbaum}(2015)}]{greenbaumIntroductionQuantumGate2015}%
  \BibitemOpen
  \bibfield  {author} {\bibinfo {author} {\bibfnamefont {Daniel}\ \bibnamefont
  {Greenbaum}},\ }\bibfield  {title} {\enquote {\bibinfo {title} {Introduction
  to {{Quantum Gate Set Tomography}}},}\ }\href
  {http://arxiv.org/abs/1509.02921} {\bibfield  {journal} {\bibinfo  {journal}
  {arXiv:1509.02921 [quant-ph]}\ } (\bibinfo {year} {2015})}\BibitemShut
  {NoStop}%
\bibitem [{Note1()}]{Note1}%
  \BibitemOpen
  \bibinfo {note} {Note that here we have abused the notation of $\protect
  \mathcal {R} \protect \mathcal {N}$ assuming it will only act on the logical
  Pauli basis $\protect \{\protect \pket {\protect \overline {G}\Pi
  _{0}}\protect \}$ instead on all of the physical Pauli basis.}\BibitemShut
  {Stop}%
\bibitem [{\citenamefont {Geller}\ and\ \citenamefont
  {Zhou}(2013)}]{gellerEfficientErrorModels2013}%
  \BibitemOpen
  \bibfield  {author} {\bibinfo {author} {\bibfnamefont {Michael~R.}\
  \bibnamefont {Geller}}\ and\ \bibinfo {author} {\bibfnamefont {Zhongyuan}\
  \bibnamefont {Zhou}},\ }\bibfield  {title} {\enquote {\bibinfo {title}
  {Efficient error models for fault-tolerant architectures and the {{Pauli}}
  twirling approximation},}\ }\href {\doibase 10.1103/PhysRevA.88.012314}
  {\bibfield  {journal} {\bibinfo  {journal} {Physical Review A}\ }\textbf
  {\bibinfo {volume} {88}} (\bibinfo {year} {2013}),\
  10.1103/PhysRevA.88.012314}\BibitemShut {NoStop}%
\bibitem [{\citenamefont {Cai}\ and\ \citenamefont
  {Benjamin}(2019)}]{caiConstructingSmallerPauli2019}%
  \BibitemOpen
  \bibfield  {author} {\bibinfo {author} {\bibfnamefont {Zhenyu}\ \bibnamefont
  {Cai}}\ and\ \bibinfo {author} {\bibfnamefont {Simon~C.}\ \bibnamefont
  {Benjamin}},\ }\bibfield  {title} {\enquote {\bibinfo {title} {Constructing
  {{Smaller Pauli Twirling Sets}} for {{Arbitrary Error Channels}}},}\ }\href
  {\doibase 10.1038/s41598-019-46722-7} {\bibfield  {journal} {\bibinfo
  {journal} {Scientific Reports}\ }\textbf {\bibinfo {volume} {9}},\ \bibinfo
  {pages} {1--11} (\bibinfo {year} {2019})}\BibitemShut {NoStop}%
\bibitem [{Note2()}]{Note2}%
  \BibitemOpen
  \bibinfo {note} {More precisely, we can find at least one physical
  representation of logical $\protect \overline {G}$ out of all of its
  logically equivalent counter-parts that satisfy this symmetry
  condition.}\BibitemShut {Stop}%
\bibitem [{\citenamefont {Wallman}\ and\ \citenamefont
  {Emerson}(2016)}]{wallmanNoiseTailoringScalable2016}%
  \BibitemOpen
  \bibfield  {author} {\bibinfo {author} {\bibfnamefont {Joel~J.}\ \bibnamefont
  {Wallman}}\ and\ \bibinfo {author} {\bibfnamefont {Joseph}\ \bibnamefont
  {Emerson}},\ }\bibfield  {title} {\enquote {\bibinfo {title} {Noise tailoring
  for scalable quantum computation via randomized compiling},}\ }\href
  {\doibase 10.1103/PhysRevA.94.052325} {\bibfield  {journal} {\bibinfo
  {journal} {Physical Review A}\ }\textbf {\bibinfo {volume} {94}},\ \bibinfo
  {pages} {052325} (\bibinfo {year} {2016})}\BibitemShut {NoStop}%
\bibitem [{\citenamefont {Rahn}\ \emph {et~al.}(2002)\citenamefont {Rahn},
  \citenamefont {Doherty},\ and\ \citenamefont
  {Mabuchi}}]{rahnExactPerformanceConcatenated2002}%
  \BibitemOpen
  \bibfield  {author} {\bibinfo {author} {\bibfnamefont {Benjamin}\
  \bibnamefont {Rahn}}, \bibinfo {author} {\bibfnamefont {Andrew~C.}\
  \bibnamefont {Doherty}}, \ and\ \bibinfo {author} {\bibfnamefont {Hideo}\
  \bibnamefont {Mabuchi}},\ }\bibfield  {title} {\enquote {\bibinfo {title}
  {Exact performance of concatenated quantum codes},}\ }\href {\doibase
  10.1103/PhysRevA.66.032304} {\bibfield  {journal} {\bibinfo  {journal}
  {Physical Review A}\ }\textbf {\bibinfo {volume} {66}},\ \bibinfo {pages}
  {032304} (\bibinfo {year} {2002})}\BibitemShut {NoStop}%
\bibitem [{\citenamefont {Guti{\'e}rrez}\ \emph {et~al.}(2016)\citenamefont
  {Guti{\'e}rrez}, \citenamefont {Smith}, \citenamefont {Lulushi},
  \citenamefont {Janardan},\ and\ \citenamefont
  {Brown}}]{gutierrezErrorsPseudothresholdsIncoherent2016}%
  \BibitemOpen
  \bibfield  {author} {\bibinfo {author} {\bibfnamefont {Mauricio}\
  \bibnamefont {Guti{\'e}rrez}}, \bibinfo {author} {\bibfnamefont {Conor}\
  \bibnamefont {Smith}}, \bibinfo {author} {\bibfnamefont {Livia}\ \bibnamefont
  {Lulushi}}, \bibinfo {author} {\bibfnamefont {Smitha}\ \bibnamefont
  {Janardan}}, \ and\ \bibinfo {author} {\bibfnamefont {Kenneth~R.}\
  \bibnamefont {Brown}},\ }\bibfield  {title} {\enquote {\bibinfo {title}
  {Errors and pseudothresholds for incoherent and coherent noise},}\ }\href
  {\doibase 10.1103/PhysRevA.94.042338} {\bibfield  {journal} {\bibinfo
  {journal} {Physical Review A}\ }\textbf {\bibinfo {volume} {94}},\ \bibinfo
  {pages} {042338} (\bibinfo {year} {2016})}\BibitemShut {NoStop}%
\bibitem [{\citenamefont {Ng}\ \emph {et~al.}(2011)\citenamefont {Ng},
  \citenamefont {Lidar},\ and\ \citenamefont
  {Preskill}}]{ngCombiningDynamicalDecoupling2011}%
  \BibitemOpen
  \bibfield  {author} {\bibinfo {author} {\bibfnamefont {Hui~Khoon}\
  \bibnamefont {Ng}}, \bibinfo {author} {\bibfnamefont {Daniel~A.}\
  \bibnamefont {Lidar}}, \ and\ \bibinfo {author} {\bibfnamefont {John}\
  \bibnamefont {Preskill}},\ }\bibfield  {title} {\enquote {\bibinfo {title}
  {Combining dynamical decoupling with fault-tolerant quantum computation},}\
  }\href {\doibase 10.1103/PhysRevA.84.012305} {\bibfield  {journal} {\bibinfo
  {journal} {Physical Review A}\ }\textbf {\bibinfo {volume} {84}},\ \bibinfo
  {pages} {012305} (\bibinfo {year} {2011})}\BibitemShut {NoStop}%
\bibitem [{\citenamefont {Viola}\ \emph {et~al.}(1999)\citenamefont {Viola},
  \citenamefont {Knill},\ and\ \citenamefont
  {Lloyd}}]{violaDynamicalDecouplingOpen1999}%
  \BibitemOpen
  \bibfield  {author} {\bibinfo {author} {\bibfnamefont {Lorenza}\ \bibnamefont
  {Viola}}, \bibinfo {author} {\bibfnamefont {Emanuel}\ \bibnamefont {Knill}},
  \ and\ \bibinfo {author} {\bibfnamefont {Seth}\ \bibnamefont {Lloyd}},\
  }\bibfield  {title} {\enquote {\bibinfo {title} {Dynamical {{Decoupling}} of
  {{Open Quantum Systems}}},}\ }\href {\doibase 10.1103/PhysRevLett.82.2417}
  {\bibfield  {journal} {\bibinfo  {journal} {Physical Review Letters}\
  }\textbf {\bibinfo {volume} {82}},\ \bibinfo {pages} {2417--2421} (\bibinfo
  {year} {1999})}\BibitemShut {NoStop}%
\bibitem [{\citenamefont {Zanardi}(1999)}]{zanardiSymmetrizingEvolutions1999}%
  \BibitemOpen
  \bibfield  {author} {\bibinfo {author} {\bibfnamefont {Paolo}\ \bibnamefont
  {Zanardi}},\ }\bibfield  {title} {\enquote {\bibinfo {title} {Symmetrizing
  evolutions},}\ }\href {\doibase 10.1016/S0375-9601(99)00365-5} {\bibfield
  {journal} {\bibinfo  {journal} {Physics Letters A}\ }\textbf {\bibinfo
  {volume} {258}},\ \bibinfo {pages} {77--82} (\bibinfo {year}
  {1999})}\BibitemShut {NoStop}%
\bibitem [{\citenamefont {Uhrig}(2007)}]{uhrigKeepingQuantumBit2007}%
  \BibitemOpen
  \bibfield  {author} {\bibinfo {author} {\bibfnamefont {G{\"o}tz~S.}\
  \bibnamefont {Uhrig}},\ }\bibfield  {title} {\enquote {\bibinfo {title}
  {Keeping a quantum bit alive by optimized {$\pi$}-pulse sequences},}\ }\href
  {\doibase 10.1103/PhysRevLett.98.100504} {\bibfield  {journal} {\bibinfo
  {journal} {Physical Review Letters}\ }\textbf {\bibinfo {volume} {98}},\
  \bibinfo {pages} {100504} (\bibinfo {year} {2007})}\BibitemShut {NoStop}%
\bibitem [{\citenamefont {Kessler}\ \emph {et~al.}(2014)\citenamefont
  {Kessler}, \citenamefont {Lovchinsky}, \citenamefont {Sushkov},\ and\
  \citenamefont {Lukin}}]{kesslerQuantumErrorCorrection2014}%
  \BibitemOpen
  \bibfield  {author} {\bibinfo {author} {\bibfnamefont {E.~M.}\ \bibnamefont
  {Kessler}}, \bibinfo {author} {\bibfnamefont {I.}~\bibnamefont {Lovchinsky}},
  \bibinfo {author} {\bibfnamefont {A.~O.}\ \bibnamefont {Sushkov}}, \ and\
  \bibinfo {author} {\bibfnamefont {M.~D.}\ \bibnamefont {Lukin}},\ }\bibfield
  {title} {\enquote {\bibinfo {title} {Quantum {{Error Correction}} for
  {{Metrology}}},}\ }\href {\doibase 10.1103/PhysRevLett.112.150802} {\bibfield
   {journal} {\bibinfo  {journal} {Physical Review Letters}\ }\textbf {\bibinfo
  {volume} {112}},\ \bibinfo {pages} {150802} (\bibinfo {year}
  {2014})}\BibitemShut {NoStop}%
\bibitem [{\citenamefont {D{\"u}r}\ \emph {et~al.}(2014)\citenamefont
  {D{\"u}r}, \citenamefont {Skotiniotis}, \citenamefont {Fr{\"o}wis},\ and\
  \citenamefont {Kraus}}]{durImprovedQuantumMetrology2014}%
  \BibitemOpen
  \bibfield  {author} {\bibinfo {author} {\bibfnamefont {W.}~\bibnamefont
  {D{\"u}r}}, \bibinfo {author} {\bibfnamefont {M.}~\bibnamefont
  {Skotiniotis}}, \bibinfo {author} {\bibfnamefont {F.}~\bibnamefont
  {Fr{\"o}wis}}, \ and\ \bibinfo {author} {\bibfnamefont {B.}~\bibnamefont
  {Kraus}},\ }\bibfield  {title} {\enquote {\bibinfo {title} {Improved
  {{Quantum Metrology Using Quantum Error Correction}}},}\ }\href {\doibase
  10.1103/PhysRevLett.112.080801} {\bibfield  {journal} {\bibinfo  {journal}
  {Physical Review Letters}\ }\textbf {\bibinfo {volume} {112}},\ \bibinfo
  {pages} {080801} (\bibinfo {year} {2014})}\BibitemShut {NoStop}%
\bibitem [{\citenamefont {Zhou}\ \emph {et~al.}(2018)\citenamefont {Zhou},
  \citenamefont {Zhang}, \citenamefont {Preskill},\ and\ \citenamefont
  {Jiang}}]{zhouAchievingHeisenbergLimit2018}%
  \BibitemOpen
  \bibfield  {author} {\bibinfo {author} {\bibfnamefont {Sisi}\ \bibnamefont
  {Zhou}}, \bibinfo {author} {\bibfnamefont {Mengzhen}\ \bibnamefont {Zhang}},
  \bibinfo {author} {\bibfnamefont {John}\ \bibnamefont {Preskill}}, \ and\
  \bibinfo {author} {\bibfnamefont {Liang}\ \bibnamefont {Jiang}},\ }\bibfield
  {title} {\enquote {\bibinfo {title} {Achieving the {{Heisenberg}} limit in
  quantum metrology using quantum error correction},}\ }\href {\doibase
  10.1038/s41467-017-02510-3} {\bibfield  {journal} {\bibinfo  {journal}
  {Nature Communications}\ }\textbf {\bibinfo {volume} {9}},\ \bibinfo {pages}
  {78} (\bibinfo {year} {2018})}\BibitemShut {NoStop}%
\bibitem [{\citenamefont {{Bonet-Monroig}}\ \emph {et~al.}(2018)\citenamefont
  {{Bonet-Monroig}}, \citenamefont {Sagastizabal}, \citenamefont {Singh},\ and\
  \citenamefont {O'Brien}}]{bonet-monroigLowcostErrorMitigation2018a}%
  \BibitemOpen
  \bibfield  {author} {\bibinfo {author} {\bibfnamefont {X.}~\bibnamefont
  {{Bonet-Monroig}}}, \bibinfo {author} {\bibfnamefont {R.}~\bibnamefont
  {Sagastizabal}}, \bibinfo {author} {\bibfnamefont {M.}~\bibnamefont {Singh}},
  \ and\ \bibinfo {author} {\bibfnamefont {T.~E.}\ \bibnamefont {O'Brien}},\
  }\bibfield  {title} {\enquote {\bibinfo {title} {Low-cost error mitigation by
  symmetry verification},}\ }\href {\doibase 10.1103/PhysRevA.98.062339}
  {\bibfield  {journal} {\bibinfo  {journal} {Physical Review A}\ }\textbf
  {\bibinfo {volume} {98}},\ \bibinfo {pages} {062339} (\bibinfo {year}
  {2018})}\BibitemShut {NoStop}%
\bibitem [{\citenamefont {McArdle}\ \emph {et~al.}(2019)\citenamefont
  {McArdle}, \citenamefont {Yuan},\ and\ \citenamefont
  {Benjamin}}]{mcardleErrorMitigatedDigitalQuantum2019}%
  \BibitemOpen
  \bibfield  {author} {\bibinfo {author} {\bibfnamefont {Sam}\ \bibnamefont
  {McArdle}}, \bibinfo {author} {\bibfnamefont {Xiao}\ \bibnamefont {Yuan}}, \
  and\ \bibinfo {author} {\bibfnamefont {Simon}\ \bibnamefont {Benjamin}},\
  }\bibfield  {title} {\enquote {\bibinfo {title} {Error-{{Mitigated Digital
  Quantum Simulation}}},}\ }\href {\doibase 10.1103/PhysRevLett.122.180501}
  {\bibfield  {journal} {\bibinfo  {journal} {Physical Review Letters}\
  }\textbf {\bibinfo {volume} {122}},\ \bibinfo {pages} {180501} (\bibinfo
  {year} {2019})}\BibitemShut {NoStop}%
\bibitem [{\citenamefont {McClean}\ \emph {et~al.}(2019)\citenamefont
  {McClean}, \citenamefont {Jiang}, \citenamefont {Rubin}, \citenamefont
  {Babbush},\ and\ \citenamefont {Neven}}]{mccleanDecodingQuantumErrors2019}%
  \BibitemOpen
  \bibfield  {author} {\bibinfo {author} {\bibfnamefont {Jarrod~R.}\
  \bibnamefont {McClean}}, \bibinfo {author} {\bibfnamefont {Zhang}\
  \bibnamefont {Jiang}}, \bibinfo {author} {\bibfnamefont {Nicholas~C.}\
  \bibnamefont {Rubin}}, \bibinfo {author} {\bibfnamefont {Ryan}\ \bibnamefont
  {Babbush}}, \ and\ \bibinfo {author} {\bibfnamefont {Hartmut}\ \bibnamefont
  {Neven}},\ }\bibfield  {title} {\enquote {\bibinfo {title} {Decoding quantum
  errors with subspace expansions},}\ }\href {http://arxiv.org/abs/1903.05786}
  {\bibfield  {journal} {\bibinfo  {journal} {arXiv:1903.05786 [physics,
  physics:quant-ph]}\ } (\bibinfo {year} {2019})}\BibitemShut {NoStop}%
\bibitem [{\citenamefont {Kimmel}\ \emph {et~al.}(2014)\citenamefont {Kimmel},
  \citenamefont {{da Silva}}, \citenamefont {Ryan}, \citenamefont {Johnson},\
  and\ \citenamefont {Ohki}}]{kimmelRobustExtractionTomographic2014}%
  \BibitemOpen
  \bibfield  {author} {\bibinfo {author} {\bibfnamefont {Shelby}\ \bibnamefont
  {Kimmel}}, \bibinfo {author} {\bibfnamefont {Marcus~P.}\ \bibnamefont {{da
  Silva}}}, \bibinfo {author} {\bibfnamefont {Colm~A.}\ \bibnamefont {Ryan}},
  \bibinfo {author} {\bibfnamefont {Blake~R.}\ \bibnamefont {Johnson}}, \ and\
  \bibinfo {author} {\bibfnamefont {Thomas}\ \bibnamefont {Ohki}},\ }\bibfield
  {title} {\enquote {\bibinfo {title} {Robust {{Extraction}} of {{Tomographic
  Information}} via {{Randomized Benchmarking}}},}\ }\href {\doibase
  10.1103/PhysRevX.4.011050} {\bibfield  {journal} {\bibinfo  {journal}
  {Physical Review X}\ }\textbf {\bibinfo {volume} {4}},\ \bibinfo {pages}
  {011050} (\bibinfo {year} {2014})}\BibitemShut {NoStop}%
\bibitem [{\citenamefont {Helsen}\ \emph {et~al.}(2019)\citenamefont {Helsen},
  \citenamefont {Battistel},\ and\ \citenamefont
  {Terhal}}]{helsenSpectralQuantumTomography2019}%
  \BibitemOpen
  \bibfield  {author} {\bibinfo {author} {\bibfnamefont {Jonas}\ \bibnamefont
  {Helsen}}, \bibinfo {author} {\bibfnamefont {Francesco}\ \bibnamefont
  {Battistel}}, \ and\ \bibinfo {author} {\bibfnamefont {Barbara~M.}\
  \bibnamefont {Terhal}},\ }\bibfield  {title} {\enquote {\bibinfo {title}
  {Spectral {{Quantum Tomography}}},}\ }\href
  {https://arxiv.org/abs/1904.00177v1} {\  (\bibinfo {year}
  {2019})}\BibitemShut {NoStop}%
\end{thebibliography}
    %merlin.mbs apsrev4-1.bst 2010-07-25 4.21a (PWD, AO, DPC) hacked
%Control: key (0)
%Control: author (0) dotless jnrlst
%Control: editor formatted (1) identically to author
%Control: production of article title (0) allowed
%Control: page (1) range
%Control: year (0) verbatim
%Control: production of eprint (0) enabled
%

\end{document}